\documentclass[11pt]{article}
\usepackage{qic}
\usepackage{latexsym}
\usepackage{amsmath}
\usepackage{amsfonts}
\usepackage{graphicx}
\usepackage{url}
\usepackage{epsfig}
\usepackage{eepic}

\def\bbC{\mathbb{C}}

\newcommand{\ba}{\begin{array}}
\newcommand{\ea}{\end{array}}

\newcommand{\secref}[1]{Section~$\ref{sec:#1}$}

\def\eps{\epsilon}

\def\be{\begin{equation}}
\def\ee{\end{equation}}
\def\bea{\begin{eqnarray}}
\def\eea{\end{eqnarray}}
\newcommand{\ket}[1]{|#1\rangle}
\newcommand{\bra}[1]{\langle #1|}

\newcommand{\ketbra}[1]{|#1\rangle\langle #1|}

\newcommand{\thmref}[1]{Theorem~\ref{thm:#1}}

\newtheorem{defi}[theorem]{Definition}

\newcommand{\ignore}[1]{}

\begin{document}

\title{The complexity of quantum spin systems on a two-dimensional square
lattice}
\author{Roberto Oliveira \thanks{IBM Watson Research Center, Yorktown Heights, NY, USA 10598. \texttt{riolivei@us.ibm.com} }\and
Barbara M. Terhal \thanks{IBM Watson Research Center, Yorktown
Heights, NY, USA 10598. \texttt{terhal@watson.ibm.com} }}
 \maketitle

\begin{abstract}
The problem {\rm \small 2-LOCAL HAMILTONIAN} has been shown to be
complete for the quantum computational class QMA \cite{KKR:hamsiam}.
In this paper we show that this important problem remains
QMA-complete when the interactions of the 2-local Hamiltonian are
between qubits on a two-dimensional (2-D) square lattice.
Our results are partially derived with novel perturbation gadgets that employ
mediator qubits which allow us to manipulate $k$-local interactions.
As a side result, we obtain that quantum adiabatic computation using
2-local interactions restricted to a 2-D square lattice is
equivalent to the circuit model of quantum computation. Our perturbation
method also shows how any stabilizer space associated with
a $k$-local stabilizer (for constant $k$) can be generated as an approximate ground-space of
a 2-local Hamiltonian.
\end{abstract}


\section{Introduction}
The novel possibilities that quantum mechanics brings to
information processing have been the subject of intense study in
recent years. In particular, much interest has been devoted to
understanding the strengths and weaknesses of quantum computing as
it pertains to important problems in computer science and physics.

An important part of this research program consists of
understanding which families of quantum systems are {\em
computationally complex}. This complexity can manifest itself in
two ways. On the one hand, a positive result shows that a given
family of systems is ``complicated enough" to efficiently
implement universal quantum computation. On the other hand, a
negative result shows that certain questions about such systems
are unlikely to be efficiently answerable. A proof of {\rm
QMA}-completeness offers compelling evidence of the negative kind
while also locating the given problem in the complexity hierarchy,
since QMA, --the class of decision problems that can be
efficiently solved on a quantum computer with access to a quantum
witness--, is analogous to the classical complexity classes NP and
MA. More precisely, the class {\rm QMA} is defined as

\begin{defi}[QMA]
A {\em promise problem} $L = L_{yes}\cup L_{no}\subseteq \{0,1\}^*$
is in {\rm QMA} if there is an efficient (of ${\rm poly}(|x|)$ size) uniform quantum circuit family
$\{V_x\}_{x\in\{0,1\}^*}$ such that
\bea
\forall\, x\in L_{yes},\;\; \exists\, \ket{\psi_x} \in {\cal H}^{\otimes {\rm poly}(|x|)}, & {\rm Prob}(V_x(\ketbra{\psi_x})=1) \geq 2/3,
\nonumber
\eea
and
\bea
\forall\, x\in L_{no},\;\; \forall\, \ket{\xi} \in {\cal H}^{\otimes {\rm poly}(|x|)}, & {\rm Prob}(V_x(\ketbra{\xi})=1) \leq 1/3.
\nonumber
\eea
\end{defi}

The work on finding QMA-complete problems was jump-started by a
`quantum Cook-Levin Theorem' proved by Kitaev \cite{KSV:computation}
(see also the survey \cite{AN:qnp}). Kitaev showed that the promise
problem {\small $k$-LOCAL HAMILTONIAN} for $k=5$ is QMA-complete.
Before we state this problem, let us review some definitions. A
Hamiltonian is a Hermitian operator. A Hamiltonian on $n$ qubits is
$k$-local for constant $k$ if it can be written as $\sum_{j=1}^r H_j$ where each term
$H_j$ acts non-trivially on at most $k$ qubits and thus $r \leq {\rm poly}(n)$.
Furthermore, we require that $||H_j|| \leq {\rm poly}(n)$ and the
entries of $H_j$ are specified by ${\rm poly}(n)$ bits.  The
smallest eigenvalue of $H$, sometimes called the `ground state
energy' of $H$, will be denoted as $\lambda(H)$.

With these definitions in place one can define the promise problem
{\small $k$-LOCAL HAMILTONIAN} as:

\begin{defi}[{\rm \small $k$-LOCAL HAMILTONIAN}]
Given is a $k$-local Hamiltonian $H$ and $\alpha, \beta$ such that
$\beta-\alpha \geq \frac{1}{{\rm poly}(n)}$.  We have a promise that
either $\lambda(H) \leq \alpha$ or $\lambda(H)> \beta$. The problem
is to decide whether $\lambda(H) \leq \alpha$. When $\lambda(H) \leq
\alpha$ we say we have a `YES-instance'.
\end{defi}

Kitaev's result was strengthened in Ref.~\cite{KR:localham}, which
showed that {\small 3-LOCAL HAMILTONIAN} was QMA-complete. The
subsequent \cite{KKR:hamsiam} proved that also {\small 2-LOCAL
HAMILTONIAN} is QMA-complete.

In another direction it was first shown by Aharonov {\em et al.}
\cite{ADLLKR:adia} that adiabatic quantum computation using 3-local
Hamiltonians is computationally equivalent to quantum computation in
the circuit model. In the adiabatic computation paradigm one starts
the computation in the ground-state, i.e. the eigenstate with
smallest eigenvalue, of some Hamiltonian $H(t=0)$. The computation
proceeds by slowly (at a rate at most ${\rm poly}(n)$) changing the
parameters of the Hamiltonian $H(t)$. The adiabatic theorem (see
Ref.~\cite{AR:adia} for an accessible proof thereof) states
essentially that if the instantaneous Hamiltonian $H(t)$ has a
sufficiently large spectral gap, -- i.e. the difference between the
second smallest eigenvalue and the smallest eigenvalue is
$\Omega(1/{\rm poly}(n))$--, then the state at time $t$ during the
evolution is close to the ground-state of the instantaneous
Hamiltonian $H(t)$. At the end of the computation ($t=T$), one
measures the qubits in the ground-state of the final Hamiltonian
$H(T)$. Ref.~\cite{KKR:hamsiam} improved on the result by Aharonov
{\em et al.} by showing that any efficient quantum computation can
be efficiently simulated by an adiabatic computation employing only
2-local Hamiltonians.

These results on the complexity of Hamiltonians can be viewed as the
first (see also Ref.~\cite{JWB:qma}) in a field that is still
largely unexplored as compared to the classical case. The class of
Hamiltonian problems is likely to be a very important class of
problems in QMA. Hamiltonians govern the dynamics of quantum systems
and as such contain all the physically important information about a
quantum system. The problem of determining properties of the
spectrum, in particular the ground state (energy) or the low-lying
excitations, is a well-known problem for which a variety of methods,
both numerical and analytical, (see e.g.
\cite{nachtergaele:qspin,book:plischke&bergersen}) have been
developed. Furthermore, finding QMA-complete problems may help us in
finding new problems that are in BQP.

 Let us briefly review the classical situation. In
some sense the {\small 2-LOCAL HAMILTONIAN} problem is similar to
the {\small MAX-2-SAT} problem \cite{book:papadimitriou}. But
perhaps a better analogue is the set of problems defined with
`classical' Hamiltonians such as {\small ISING SPIN GLASS}:


\begin{defi}[{\rm \small ISING SPIN GLASS}]
Given is an interaction graph $G=(V,E)$ with Hamiltonian \be
H_G=\sum_{i,j \in E} J_{ij} \,Z_i \otimes Z_j+\sum_{i\in V} \Gamma_i
Z_i. \ee Here the couplings $J_{ij} \in \{-1,0,1\}$ and $\Gamma_i
\in \{-1,0,1\}$  and $Z=\ketbra{0}-\ketbra{1}$ is the Pauli Z
operator. The problem is to decide whether $\lambda(H_G) \leq
\alpha$ for a given $\alpha$.
\end{defi}

It is known that the problem {\small ISING SPIN GLASS}, which is a
special case of the 2-local Hamiltonian problem, is NP-complete on a
planar graph. In fact, it is even NP-complete on a planar graph when
$J_{ij}=J=1$ and $\Gamma_i=\Gamma=1$ \cite{barahona:np}. In this
paper we prove some results on the complexity of a quantum version
of this model, a quantum spin glass. Our results are based on two
ideas. The first one is a small modification to the `quantum
Cook-Levin' circuit-to-5-local Hamiltonian construction that will
prove QMA-completeness of a 5-local Hamiltonian on a `spatially
sparse' hypergraph (to be defined below). Such QMA-completeness
result on a spatially sparse hypergraph could also have been
obtained from the 6-dim particle Hamiltonian on a 2D lattice that
was constructed in \cite{ADLLKR:adia}.

Secondly, we introduce a set of mediator qubit
gadgets\footnote{These gadgets are inspired by the idea of
superexchange between particles with spin. Loosely speaking,
superexchange is the creation of an effective spin `exchange'
interaction due to a mediating particle, first calculated by H.A.
Kramers in 1934 \cite{kramers:pert}.} to manipulate $k$-local
interactions. These gadgets can be used to reduce any $k$-local
interaction for constant $k$ to a 2-local interaction. Then we use
the gadgets to reduce a 2-local Hamiltonian on a spatially sparse
graph to a 2-local Hamiltonian on a planar graph, or alternatively
to a 2-local Hamiltonian on a 2D lattice. The general technique is
based on the idea of perturbation gadgets introduced in Ref.~\cite{KKR:hamsiam}.
However the gadgets that we introduce here are
more general and more powerful than the one in Ref.~\cite{KKR:hamsiam}.

Before we state the results, let us give a few more useful
definitions. With a 2-local Hamiltonian $H_G$ acting on $n$ qubits
we can associate an {\em interaction graph} $G=(V,E)$ with $|V|=n$.
For every edge in $e \in E$ between vertices $a$ and $b$ there is a
nonzero 2-local term $H_{e}$ on qubits $a$ and $b$ such that $H_{e}$
is not 1-local nor proportional to the identity operator I. We can
write $H_G=\sum_{e \in E} H_e+\sum_{v \in V} H_v$ where $H_v$ is a
potential 1-local term on the vertex $v$. Similarly, with a
$k$-local Hamiltonian one can associate an interaction hypergraph in
which the $k$-local terms correspond to hyper-edges in which $k$
vertices are involved. We also use the following definition of a
spatially sparse hypergraph. A spatially sparse interaction
(hyper)graph $G$ is defined as a (hyper)graph in which (i) every
vertex participates in $O(1)$ hyper-edges, (ii) there is a
straight-line drawing in the plane such that every hyper-edge
overlaps with $O(1)$ other hyper-edges and the surface covered by
every hyper-edge is $O(1)$.

\indent A Pauli edge of an interaction graph $G$ is an edge between
vertices $a$ and $b$ associated with an operator $\alpha_{ab} P_a
\otimes P_b$ where $P_a,P_b$ are Pauli matrices
$X=\ket{0}\bra{1}+\ket{1}\bra{0}$,
$Y=-i\ket{0}\bra{1}+i\ket{1}\bra{0}$, $Z=\ketbra{0}-\ketbra{1}$ and
$\alpha_{ab}$ is some real number. For an interaction graph in which
every edge is a {\em Pauli edge}, the degree of a vertex is called
its {\em Pauli degree}. For such a graph, the $X$- (resp. $Y$-,
resp. $Z$-) degree of $a$ vertex a is the number of edges with
endpoint $a$ for which $P_a=X$ (resp. $P_a=Y$, resp. $P_a=Z$).

\indent We will prove the following results. First we show that

\begin{theorem}
{\rm \small 2-LOCAL HAMILTONIAN} on a planar graph with maximum
Pauli degree equal to 3 is {\rm QMA}-complete.
\label{theo:degree4}
\end{theorem}

With only a little more work, we prove that

\begin{theorem}
\mbox{\rm \small 2-LOCAL HAMILTONIAN} with Pauli interactions on a
subgraph of the 2-D square lattice is {\rm QMA}-complete.
\label{theo:squarelat}
\end{theorem}


Lastly, we answer an open problem in Ref.~\cite{ADLLKR:adia} (see
Section \ref{sec:adiabatic} for a more detailed statement of the
result), namely that:


\begin{theorem}\label{thm:adiabatic}Universal
quantum computation can be efficiently simulated by a quantum
adiabatic evolution of qubits interacting on a 2-D square
lattice.\end{theorem}

We believe that our Theorem \ref{theo:squarelat} is in some sense
the strongest result that one can expect for qubits, since we
consider it unlikely that {\small 2-LOCAL HAMILTONIAN} restricted to
a linear chain of qubits is QMA-complete. A recent surprising result
in this respect is that {\small 2-LOCAL HAMILTONIAN} on a
one-dimensional lattice with 12-dimensional {\em qudits} is
QMA-complete \cite{AGIK:1d}. With regards to \thmref{adiabatic}, one
should note that Aharonov {\em et al.} \cite{ADLLKR:adia} had
already proven that interactions of six-dimensional particles on a
two-dimensional square lattice suffice for universal quantum
adiabatic computation. Our improvement to qubits on a
two-dimensional lattice is an application of our perturbation
gadgets to \cite{ADLLKR:adia}'s 6-dim particle construction.

We would like to draw attention to the power of the perturbative
method and in particular to the gadgets that we develop in this
paper. There are a variety of interesting states that can be defined
as the ground-states or ground-spaces of $k$-local Hamiltonians.
Prime examples are the stabilizer states where the Hamiltonian
equals $H=I-\sum_i S_i$ and $S=\{S_i\}$ is a set of commuting
stabilizer operators. The ground-space is formed by all states with
$+1$ eigenvalue with respect to the stabilizer $S$ and this space is
separated by a constant gap from the rest of the spectrum. An
example is the cluster state \cite{RB:oneway}, the toric code space
\cite{kitaev:top} or any stabilizer code space. Typically, the
stabilizer operators $S_i$ are $k$-local with $k>2$ which seems to
preclude the generation of such ground-space as the ground-space of
a {\em natural} Hamiltonian, see the arguments in Ref.~\cite{nielsen:cluster}. The perturbative gadgets introduced in this
paper show how to generate a 2-local Hamiltonian which has a
ground-space with is {\em approximately} a product of a trivial
ancilla-qubit space times the ground-space of the desired $k$-local
Hamiltonian. Thus the use of ancillas and the use of approximation
get us past the constraints derived in \cite{nielsen:cluster}. If
the original $k$-local Hamiltonian has some restricted spatial
structure, one can show that the resulting 2-local Hamiltonian can
be defined on a planar graph or, if desired, on a 2-D lattice.

In the Appendix of this paper we prove a stronger perturbation theorem than what has been shown in
\cite{KKR:hamsiam}. The results in the Appendix show that under the appropriate conditions
the perturbative method does not only reproduce the eigenvalues of the target Hamiltonian,
but also the eigenstates, possibly restricted to the low-lying levels of the target Hamiltonian.
We believe that these results may have applications beyond reductions in QMA and
the adiabatic universality results in Section \ref{sec:adiabatic}.

This paper is organized as follows. In Section \ref{sec:5loc} we
show how to modify Kitaev's original 5-local Hamiltonian
construction \cite{KSV:computation} to a 5-local Hamiltonian with interactions
restricted to a spatially sparse hypergraph. In Section \ref{sec:gadgets} we introduce
our perturbation gadgets and in Section \ref{sec:medqubits} we show how to go from a 5-local to a 2-local
Hamiltonian using our basic mediator qubit gadget. In Section \ref{compgadgets} we use new variants of
the basic gadget to further reduce the 2-local Hamiltonian on a spatially sparse hypergraph to a 2-local Hamiltonian
on a planar graph of Pauli degree at most 3, Theorem
\ref{theo:degree4}. With a bit more work we reduce it to a
2-local Hamiltonian on a 2-D square lattice, Theorem
\ref{theo:squarelat}. Finally, \secref{adiabatic} presents the proof that adiabatic
quantum computation using 2-local Hamiltonians on a 2D lattice is computationally universal (\thmref{adiabatic}).

\section{A Spatially Sparse $5$-local Hamiltonian Problem}\label{sec:5loc}
We start by modifying the proof that {\small 5-LOCAL HAMILTONIAN} is
QMA-complete in Ref.~\cite{KSV:computation} (see also
\cite{AN:qnp}). The essential insight is (1) to modify any quantum
circuit to one in which any qubit is used a constant number of times
and (2) make sure that the program to execute the gates in the
correct time sequence is spatially local. We note that some of the
ideas in this section are quite similar to those behind the
adiabatic 2D-lattice Hamiltonian construction with $6$-dim particles
in Ref.~\cite{ADLLKR:adia}.

\begin{figure}[h]
\begin{center}
\includegraphics[scale=.40]{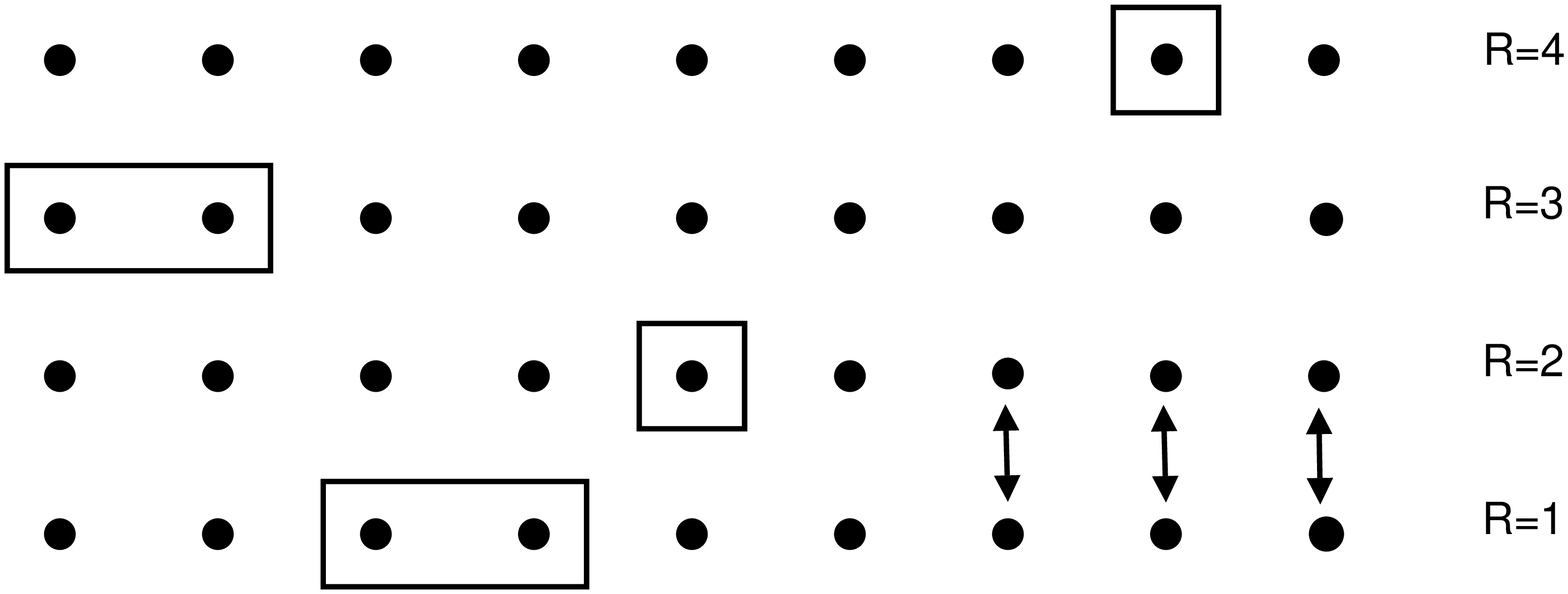}\end{center}
\fcaption{\label{fig:layout}Two-dimensional spatial layout of the
qubits in a quantum circuit for $R=4$. A qubit is indicated by a
$\bullet$. One and two-qubit gates are indicated by boxes. After the
gate is executed in row $R$, those qubits are swapped with the
qubits above them in row $R+1$. The order in which the swap and gate
operations are executed can be represented by a (time)cursor that
snakes over the circuit as follows. We start with the qubit on the
left in row $R=1$. Identity gates are applied on qubits in this row
except for the one non-trivial gate. We end up at the right and then
start swapping the qubits in row 1 with those in row 2, starting
with the qubit on the right. By doing this we end up at the left.
Now we perform a round of gate-applications (going right) on the
qubits in row $R=2$. We end up at the right and go left while
swapping the qubits in rows $R=2$ and $R=3$. We continue until all
necessary gates are executed and the computational qubits are sitting in the last row.}\end{figure}

Let a quantum circuit use $N$ qubits where $n$ qubits are input
qubits and the other $N-n$ qubits are ancilla qubits. We first
modify this circuit such that gates are executed in $R={\rm
poly}(N)$ `rounds' where in every round only $1$ (non-trivial)
gate is performed \footnote{One could do more gates per round, but
this construction is perhaps more easily explained.}. After a
round, the $N$ qubits are swapped to a next row of $N$ qubits and
then the next gate in the original circuit is executed. The total number
of qubits in this modified circuit is $M=RN$. The rows of $N$ qubits
for different rounds $R$ are depicted in Fig. \ref{fig:layout}. Let
us specify an order in which the swap and gate operations are
executed. In the first round $R=1$ we start by applying gates, $I$
and the non-trivial gate, with the qubit on the left in Figure
\ref{fig:layout}. After this round, the swapping starts with the
qubit on the right. Then again the $R=2$ gate-round starts with
qubits on the left etc. If we label the gates (including $I$) with a
time-index depending on when they are executed, then it is clear
that in this model time changes in a spatially local fashion.

We also note that in our construction, each physical qubit enters a
gate at most $3$ times, twice in a swap gate, and once in a $I$ gate
or a nontrivial gate.

In the class QMA the verifier Arthur uses a verifying quantum
circuit $V_x$ for an instance $x$. We will use the fact that we can always
replace such verifying quantum circuit by a modified verifying circuit with the
properties that we derived above.

Given any instance $x$ of a promise problem $L \in QMA$ and the
verification circuit $V_x$, we will construct a 5-local Hamiltonian
$H^{(5)}$ such that
\begin{itemize}
\item if on some input $\ket{\xi,0}$ $V_x$ accepts with probability
more than $1-\epsilon$ ($x$ is a YES-instance), then $H^{(5)}$ has
an eigenvalue less than $\frac{\epsilon}{p_1(n)}$ for some
polynomial $p_1(n)$.
\item if $V_x$ accepts with probability less than $\epsilon$ then all eigenvalues of $H^{(5)}$ are larger than
$\frac{1-\epsilon-\sqrt{\epsilon}}{p_2(n)}$ for some polynomial
$p_2(n)$.
\end{itemize}
Thus we can map each promise problem in QMA onto a 5-local
Hamiltonian problem where the specific restricted form of Arthur's
verifying circuit will lead to restrictions on the interactions in
the 5-local Hamiltonian, that is, the interaction hypergraph will
be spatially sparse. In particular, when $\epsilon=O(2^{-n})$ for a $n$
qubit proof from Merlin, we obtain a Hamiltonian which obeys the
promise in Definition 2. Note that Definition 1 uses $\epsilon=1/3$
but it has been shown, see e.g. \cite{MW:error}, that one can make the
error $\epsilon=O(2^{-n})$ for a $n$ qubit proof input.

Thus, these arguments will prove that the 5-local Hamiltonian problem on a
so-called spatially sparse hypergraph is QMA-hard. Since it is also
known that {\small 5-LOCAL HAMILTONIAN} is in QMA
\cite{KSV:computation}, this proves the QMA-completeness of {\small
5-LOCAL HAMILTONIAN} on a spatially sparse hypergraph.

Let us now look at the details of mapping a QMA circuit onto a
Hamiltonian problem. Our construction is a small modification from
the standard construction by Kitaev \cite{KSV:computation}. We
define a set of clock-qubits. We use $T=(2R-1)N$ clock-qubits
labeled as $c_1 \ldots, c_T$. Time $t$ will be represented as the
state $\ket{1^t 0^{T-t}}_{c_1 \ldots c_T}$ as in Ref.~\cite{KSV:computation}. Let $U_1 \ldots U_T$ be the sequence of
operations on the computational qubits of the quantum circuit $V$,
one operation for every clock-qubit $c_1, \ldots,c_T$ . The set of
operations includes the actual gates, the $I$ operations when only
time advances and the swap gates. Let ${\rm Q_{in}}$ be the set of
$n$ qubits that contain the input $\ket{\xi}$. Let $q_{\rm out}$ be
the final qubit that is measured in the quantum circuit $V_x$. The
$5$-local Hamiltonian $H^{(5)}$ that we associate with this circuit
is as follows. $H^{(5)}=H_{\rm in}+H_{\rm out}+H_{\rm
clock}+\frac{1}{2}\sum_{t=0}^T H_{\rm evolv}(t)$ where

\bea
H_{\rm in} & = & \sum_{q \notin Q_{\rm in}} \ketbra{1}_q \otimes \ketbra{100}_{c_{t_q-1},c_{t_q},c_{t_q+1}}, \nonumber \\
H_{\rm out}& = & \ketbra{0}_{q_{\rm out}} \otimes \ketbra{1}_{c_T},\nonumber \\
H_{\rm clock}& = & \sum_{t=1}^{T-1}\ketbra{01}_{c_t,c_{t+1}}. \eea
and
\bea H_{\rm evolv}(1)& =&
\ketbra{00}_{c_1,c_2}+\ketbra{10}_{c_1,c_2}
-U_1 \otimes \ket{10}\bra{00}_{c_1,c_2}-U_1^{\dagger} \otimes \ket{00}\bra{10}_{c_1,c_2}, \nonumber \\
H_{\rm evolv}(t)& = & \ketbra{100}_{c_{t-1},c_t,c_{t+1}}+\ketbra{110}_{c_{t-1},c_t, c_{t+1}} \nonumber \\
& & -U_t\otimes \ket{110}\bra{100}_{c_{t-1},c_t,c_{t+1}}-U_t^{\dagger}\otimes \ket{100}\bra{110}_{c_{t-1},c_t,c_{t+1}} ,\;\;1 < t < T \nonumber \\
H_{\rm evolv}(T)& =& \ketbra{10}_{c_{T-1},c_T}+\ketbra{11}_{c_{T-1},c_T}
-U_T\otimes \ket{11}\bra{10}_{c_{T-1},c_T}-U_T^{\dagger}\otimes \ket{10}\bra{11}_{c_{T-1},c_T}.
\eea

$H_{\rm in}$ is the only term that is different from the $5$-local
Hamiltonian considered in Ref.~\cite{KSV:computation}; it uses the
definition of a set of special times $t_q$. Before we define these
times, let us look more closely at the interactions in the
Hamiltonian and how the qubits can be laid out so that each qubit
only interacts with a set of qubits in its neighborhood. The precise
form of this neighborhood is irrelevant, we only require that the
interaction hyper-graph of this Hamiltonian spatially sparse, as
defined in the Introduction.

Given the lay-out of the computational (non-clock) qubits in Figure
\ref{fig:layout} we can `drape a string' of clock qubits over the
line following the sequence of computational steps. This ensures that
the terms in $H_{\rm evolv}$ involve qubits that are in each other's
local neighborhood. We can also ensure this locality property of
$H_{\rm out}$ by choosing the output qubit $q_{\rm out}$ to be the
last qubit on the right in the final row. Now let us consider
$H_{\rm in}$. For every qubit in the layout in Figure
\ref{fig:layout} there is a time in which the running cursor which
snakes over the circuit first arrives at this qubit. For the qubits
in $R=1$, this is when the cursor comes from the left doing the I
operations or the non-trivial gate. For the qubits in the other rows
$R> 1$, it is when the cursor, coming from the right, starts
swapping the qubit with the previous row $R-1$. These cursor actions
are represented in $H_{\rm evolv}$. For a qubit $q$ we define the
clock-qubit $c_{t_q}$ as the clock-qubit whose bit is flipped in the
interaction representing the {\em earliest} gate (the action of the
cursor) on the qubit $q$ in $H_{\rm evolv}$. Then it is clear that
the clock-qubit $c_{t_q}$ is local to the qubit $q$ and therefore
$H_{\rm in}$ again represents an interaction between qubits that are
in each other's local neighborhood. It is also clear that the role
of $H_{\rm in}$ is to make sure that the state of the qubits is set
to 0 before the gates actually act on these qubits. Note that we set
the state of all qubits (except those in $Q_{\rm in}$) to zero, also
the ones in the later rows that are merely used as dummy qubits to be
used in swaps. This is not absolutely necessary but merely
convenient.

These arguments show that the interaction hypergraph of the
Hamiltonian is spatially sparse. Note also that given a quantum
circuit with $N$ qubits one can efficiently construct the
interaction hypergraph of the corresponding Hamiltonian and draw
this hypergraph in the plane where hyperedges involving 5 qubits are
represented as five-sided polygons.

The proof of the following Lemma is analogous to the proof of Theorem 14.3 in \cite{KSV:computation}.

\begin{lemma}\label{lem:3loc}
Let
$\ket{\psi}=\sqrt{\frac{1}{T+1}}\sum_{t=0}^T\ket{\xi_t}_{q_1\dots
q_{M}}\ket{1^t 0^{T-t}}_{c_1\dots c_T}$ where $\ket{\xi_{t}} =
U_t\ket{\xi_{t-1}}$ for all $1\leq t\leq T$ and $\ket{\xi_0} =
\ket{\xi}\ket{0^{M-n}}$ for some state $\ket{\xi}$ of the input
qubits. If Arthur's verifying quantum circuit $V_x$ accepts with
probability more than $1-\epsilon$ on some input $\ket{\xi,00 \ldots 0}$
then $\bra{\psi} H^{(5)} \ket{\psi} < \frac{\epsilon}{T+1}$. If $V_x$ accepts with
probability less than $\epsilon$ on all inputs $\ket{\xi,0}$ then
all eigenvalues of $H^{(5)}$ are larger than or equal to $\frac{c(1-\epsilon-\sqrt{\epsilon})}{T^3}$
for some constant $c$.
\end{lemma}

\proof{Consider first $\bra{\psi} H^{(5)}\ket{\psi}$. We only need to check
that $\bra{\psi} H_{\rm in} \ket{\psi}=0$ since this term is
different than the one in Ref.~\cite{KSV:computation}. We note that
$H_{\rm in} \ket{\psi} \propto \sum_{q \notin Q_{\rm in}}
\ketbra{1}_q \ket{\xi_{t_q-1},1^{t_q-1} 0^{T-(t_q-1)}}=0$ since in
$\ket{\psi}$ all computational qubits are set to 0 before they are
being acted upon, i.e. qubit $q$ is the state 0 at all times $t <
t_q$. Thus $\ket{\psi}$ has zero eigenvalue with respect to all
terms in $H^{(5)}$ except $H_{\rm out}$. If $V_x$ accepts with
probability more than $1-\epsilon$, this implies that $\bra{\psi}
H^{(5)} \ket{\psi}=\bra{\psi} H_{\rm out} \ket{\psi} <
\frac{\epsilon}{T+1}$. The second part of the proof is to show that
if $V_x$ accepts with small probability, the eigenvalues of $H$ are
bounded from below. Again the proof is identical in structure to the
proof in \cite{KSV:computation} except for $H_{\rm in}$. We first
note that $H^{(5)}$ preserves the space of `legal' clock-states
${\cal S}$, i.e. clock-states of the form $\ket{1^t 0^{T-t}}$ and
thus we can consider the minimum eigenvalue problem of $H^{(5)}$ on
${\cal S}$ and ${\cal S}^{\perp}$ separately. On ${\cal S}^{\perp}$
this minimum eigenvalue is 1 since at least one of the constraints
of $H_{\rm clock}$ is not satisfied. Now we consider $H^{(5)}|_{\cal
S}$ which we can express using the definition $\ket{t} \equiv
\ket{1^t 0^{T-t}}$. We have $H_{\rm in}|_{\cal S}=\sum_{q \notin
Q_{\rm in}} \ketbra{1}_q \otimes \ketbra{t_q-1}$. As in the standard
proof we perform a rotation $W$ to a more convenient basis where
$W=\sum_{t=0}^T U_t \ldots U_1 \otimes \ket{t}\bra{t}$. Let \be H_2
\equiv W^{\dagger} H_{\rm evolv} |_{\cal S} W=I \otimes E, \ee where
$E$ is defined below Eq.~(14.9) in \cite{KSV:computation}. Let \be
H_1\equiv W^{\dagger} (H_{\rm in}+H_{\rm out})|_{\cal S} W=\sum_{q
\notin Q_{\rm in}} \ketbra{1}_q \otimes \ketbra{t_q-1}+ U^{\dagger}
\ketbra{0}_{q_{\rm out}} U \otimes \ketbra{T}, \ee where $U=U_T
\ldots U_1$. Note that $H_{\rm in}|_{\cal S}$ is unchanged by the rotation $W$ since there
are no gates acting on a qubit $q$ prior to the time $t_q$. Now we would like to use Lemma 14.4 in Ref.~\cite{KSV:computation} and lower-bound the smallest eigenvalue of
$H_1+H_2$. Let ${\cal L}_1$ and ${\cal L}_2$ be the non-empty
null-spaces of $H_1$ and $H_2$. Lemma 14.4 states that for such $H_1
\geq 0$ and $H_2 \geq 0$ we can bound
 $H_1+H_2 \geq  2 v  \sin^2 (\theta/2)$ where $v$ is the smallest non-zero eigenvalue of $H_1$ and $H_2$
and $\cos^2 \theta=\max_{\eta \in {\cal L}_2} \bra{\eta} {\bf
P}_{{\cal L}_1} \ket{\eta}$ where ${\bf P}_{{\cal L}_1}$ is the
projector on ${\cal L}_1$. The minimum of the smallest non-zero
eigenvalue of $H_1$ and $H_2$ is as in Ref.~\cite{KSV:computation},
namely $v \geq c T^{-2}$.

Now we show that, as in \cite{KSV:computation}, one can bound
$\sin^2 \theta \geq \frac{1-\epsilon-\sqrt{\epsilon}}{T+1}$. Putting
these results together shows that the minimum eigenvalue of
$H^{(5)}$ is at least $\frac{c(1-\epsilon-\sqrt{\epsilon})}{T^3}$
for some constant $c$, as claimed. As in Ref.~\cite{KSV:computation}
any state in ${\cal L}_2$ is of the form $\ket{\xi} \otimes
\frac{1}{\sqrt{T+1}}\sum_{t=0}^T \ket{t}$ where $\ket{\xi}$ is
arbitrary. We can also write ${\bf P}_{{\cal L}_1}=\sum_{t=0}^T P_t
\otimes \ketbra{t}$ where $P_T=U^{\dagger} \ketbra{1}_{q_{\rm out}}
U$, and $P_t=\Pi_{q \notin Q_{\rm in}|t_q=t+1}\ketbra{0}_q\otimes
I_{{\rm else},t}$ where $I_{{\rm else},t}$ is the $I$ operator on
all computational qubits for which $t_q \neq t+1$. At some times
$P_t$ may just be $I$ on all qubits. Here $\Pi_{q \notin Q_{\rm
in}|t_q=t+1}$ is tensor product of $\ketbra{0}$ for all qubits
$q$ for which $t_q=t+1$. Thus we need to bound \be \cos^2
\theta=\frac{1}{T+1}\max_{\xi} \bra{\xi} \sum_t P_t \ket{\xi}. \ee
All $P_t$ for $t < T$ commute and their common eigenspace is the
space where all qubits $q \notin Q_{\rm in}$ are set to $\ket{00
\ldots 0}$. We can write any $\ket{\xi}$ as $\ket{\xi}=\alpha
\ket{00 \ldots 0, \psi_0}+\ket{\beta}$ where $\psi_0$ is a state for
all qubits in $Q_{\rm in}$ and $\ket{\beta}$ is a state with norm
$1-|\alpha|^2$ in which at least one of the $k$ non-input qubits is
not in $\ket{0}$. Thus we have \bea \cos^2 \theta \leq
\frac{1}{T+1}\left(|\alpha|^2 T+ |\alpha|^2 \bra{0,\psi_0} P_T
\ket{0,\psi_0}+ 2 |\alpha|\, |\bra{0,\psi_0} P_T \ket{\beta}|+(T-1)
\bra{\beta} \beta\rangle+\bra{\beta} P_T \ket{\beta}\right). \eea
Given the acceptance probability of the circuit $V_x$ we can bound
$\bra{0,\psi_0} P_T \ket{0,\psi_0} < \epsilon$. We also bound
$\bra{\beta} P_T \ket{\beta} \leq \bra{\beta} \beta \rangle$. This
gives \be \cos^2 \theta \leq \frac{1}{T+1}\left(T+ |\alpha|^2
\epsilon+ 2 |\alpha| \sqrt{\epsilon} \sqrt{1-|\alpha|^2}\right) \leq
1-\frac{1-\epsilon-\sqrt{\epsilon}}{T+1}. \ee
}

\section{Perturbation Theory}\label{sec:gadgets}

In this section we introduce the perturbation method. Our main new
idea is the use of mediator qubits that perturbatively generate
interactions. The mediator qubits are weakly coupled to the other
qubits and to lowest order in the perturbation this coupling
generates an interaction between the other qubits, see Section
\ref{sec:medqubits}. We will show as a first step how this can be
used to reduce any $k$-local Hamiltonian problem to a 3-local
Hamiltonian problem. We can then use the perturbation gadget in
\cite{KKR:hamsiam} to reduce a 3-local to a 2-local Hamiltonian (we
also sketch an alternative mediator qubit method). To reduce a
2-local Hamiltonian to a 2-local Hamiltonian on a 2D lattice or a
planar graph, we need a few other applications of our mediator qubit
gadgets which will be introduced in Section \ref{compgadgets}.

In Ref.~\cite{KKR:hamsiam} the authors reduce the problem
$\mbox{ \small 3-LOCAL HAMILTONIAN}$ to $\mbox{ \small 2-LOCAL
HAMILTONIAN}$ by introducing a perturbation gadget. The idea is to
approximate $\lambda(H_{\rm target})$ of a desired (3-local)
Hamiltonian $H_{\rm target}$ by $\lambda(\tilde{H})$ of a 2-local
Hamiltonian $\tilde{H}$ where $\lambda(\tilde{H})$ is calculated
using perturbation theory. One sets $\tilde{H}=H+V$ where $H$ is the
`unperturbed' Hamiltonian which has a large spectral gap $\Delta$
and $V$ is a {\em small} perturbation operator. We will choose $H$
such that it has a degenerate ground-space associated with
eigenvalue 0 and the eigenvalues of the `excited' eigenstates are at
least $\Delta$. The effect of the perturbation $V$ is to lift the
degeneracy in the ground-space and create the target Hamiltonian
in this space.

More accurately, we have a Hilbert space ${\cal L}={\cal L}_+ \oplus
{\cal L}_-$ where ${\cal L}_-$ is the ground-space of $H$. Let
$\Pi_{\pm}$ be the projectors on ${\cal L}_{\pm}$. For some operator
$X$ we define $X_{++}=\Pi_+ X \Pi_+,X_{-+}=\Pi_- X
\Pi_+,X_{+-}=\Pi_+ X \Pi_-,X_{--}=\Pi_- X \Pi_-$ and $X_{+}\equiv
X_{++}$. In order to calculate the perturbed eigenvalues, one
introduces the self-energy operator $\Sigma_-(z)$ for real-valued
$z$ \be \Sigma_-(z)=H_-+V_{--}+V_{-+}G_+(I_+-V_{++}G_+)^{-1}V_{+-},
\label{selfe} \ee where we can perturbatively expand \be
(I_+-V_{++}G_+)^{-1}=I_++V_{++}G_+ +V_{++}G_+ V_{++}G_++\ldots.
\label{pertexpand}\ee Here $G_{+}$, called the unperturbed Green's
function (or resolvent) in the physics literature, is defined by \be
G_{+}^{-1}=zI_+-H_+. \ee In Ref.~\cite{KKR:hamsiam} the following
theorem is proved (here we state the case where the ground-space of
$H$ has eigenvalue 0 and $H$ has a spectral gap $\Delta$ above the
ground-space):

\begin{theorem}(\cite{KKR:hamsiam})
Let $||V|| \leq \Delta/2$ where $\Delta$ is the spectral gap of $H$ and $\lambda(H)=0$.
Let $\tilde{H}|_{< \Delta/2}$ be the restriction of $\tilde{H}=H+V$ to the space of
eigenstates with eigenvalues less than $\Delta/2$. Let there be an
effective Hamiltonian $H_{\rm eff}$ with ${\rm Spec}(H_{\rm eff})
\subseteq [a,b]$. If the self-energy $\Sigma_-(z)$ for all $z \in
[a-\epsilon,b+\epsilon]$ where $a < b < \Delta/2-\epsilon$ for some
$\epsilon > 0$, has the property that \be ||\Sigma_-(z)-H_{\rm
eff}|| \leq \epsilon, \ee then each eigenvalue $\tilde{\lambda}_j$
of $\tilde{H}|_{< \Delta/2}$ is $\epsilon$-close to the $j$th
eigenvalue of $H_{\rm eff}$. In particular \be |\lambda(H_{\rm
eff})-\lambda(\tilde{H})| \leq \epsilon. \ee \label{kkrtheorem}
\end{theorem}

This theorem can be generalized to Theorem \ref{thm:perturbation}
proved in the Appendix. Theorem \ref{thm:perturbation} shows that
under appropriate conditions, the effective Hamiltonian is
approximately identical to ${\tilde H}$ restricted to its low-lying
eigenspaces. With the same technique we also prove Lemma
\ref{lem:groundpert} in the Appendix which shows that the
ground-space of a target Hamiltonian can be generated perturbatively
(under the assumption that the target Hamiltonian has a 1/{\rm
poly}($n$) gap). Lemma \ref{lem:groundpert} was also proved in
\cite{KKR:hamsiam} in the special case that the ground-space is
non-degenerate.

\subsection{Mediator Qubit Gadgets}
\label{sec:medqubits}

In the following explanation of the gadgets we will refer to $H_{\rm
target}$ as the desired Hamiltonian that we want to generate
perturbatively and the effective Hamiltonian is $H_{\rm eff}=H_{\rm
target} \otimes \ketbra{00 \ldots}$, i.e. the ancillary `mediator'
qubits are in their ground-state $\ket{00 \ldots 0}$.

The gadgets that we introduce below to accomplish the reduction are
what we call mediator qubit gadgets and seem to be useful in general
to manipulate $k$-local interactions. The idea is that we replace a
direct interaction between two groups of $\lceil k/2 \rceil$ qubits with {\em indirect}
interactions through a mediator qubit. In the ground-state of the
unperturbed Hamiltonian $H$ the mediator qubit is in state
$\ket{0}$. The perturbation $V$ is chosen such that interaction with
the other qubits can flip the mediator qubit. The perturbative
corrections to the self-energy, up to second order in the perturbation, involve the process of flipping the mediator qubit by
interaction with a group of qubits $a$ and flipping the mediator qubit back to
$\ket{0}$ by a second interaction with a group of qubits $b$. If $a=b$ we
potentially obtain some $\lceil k/2\rceil$-local terms. For $a \neq b$ we obtain an
effective $k$-local interaction involving groups $a$ and $b$.
This gadget could also be used with three or more groups of qubits (or higher dimensional
quantum systems); in this case interactions would be generated between all groups
of qubits. An example of such application is the Cross gadget, explained in Section \ref{compgadgets}.

\begin{figure}[h]\begin{center}\includegraphics[scale=.30]{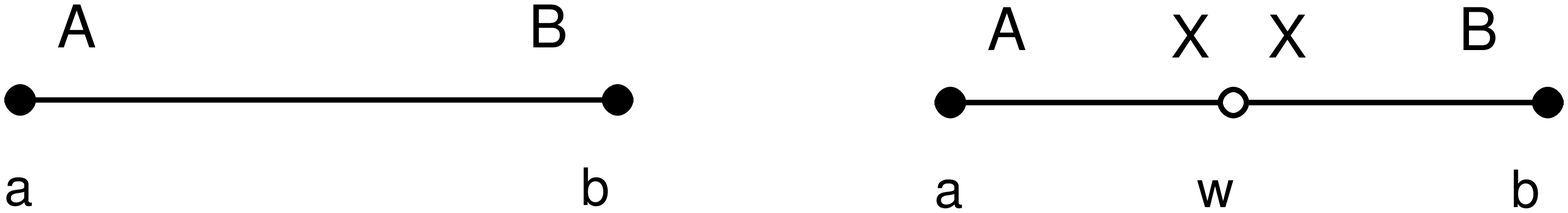}\end{center}
\fcaption{\label{fig:edgesub} Subdivision gadget. A $k$-local interaction is
reduced to $\lceil k/2 \rceil+1$-local interactions using a mediator qubit vertex $w$.
The operators $A,B,X$ next to the edges indicate
which operators correspond to the edges.}\end{figure}

{\sc Subdivision Gadget.} Assume that a $k$-local operator
associated with (hyper)edge $ab$ is of the form $A \otimes
B$ and let $r=\max(||A||,||B||)$.
The hyper-edge $ab$ is part of a larger (hyper)graph and a corresponding Hamiltonian.
Let all other terms in the Hamiltonian be $H_{\rm else}$. We can write
the Hamiltonian as \be H_{\rm target}=(H_{\rm else}+A^2/2+B^2/2)-(-A+
B)^2/2 \equiv H_{\rm else}'-(-A+B)^2/2, \ee
so that $H_{\rm else}'$ contains some additional $\lceil k/2 \rceil$-local terms as compared to $H_{\rm else}$.
W.l.o.g. we assume that $\max(||H_{\rm else}'||,r) \geq 1$.

The terms in the gadget Hamiltonian
$\tilde{H}=H+V$ are the following \be H=\Delta \ketbra{1}_w, \;\; V
=H_{\rm else}'+ \sqrt{\Delta/2} \left(-A+B\right)
\otimes X_w.\ee  The operator $X_w$
is the Pauli X operator acting on qubit $w$. The degenerate
ground-space ${\cal L}_-$ of $H$ has the mediator qubit in the state
$\ket{0}$. We have the following: $H_-=0$, $G_+(z)=
\frac{\ketbra{1}_w}{z-\Delta}$, $V_{--} = H_{\rm else}'\otimes
\ketbra{0}_w$ and \be V_{+-} = \sqrt{\Delta/2} (-A  +
B)\otimes \ket{1}\bra{0}_w. \ee

Thus the self-energy $\Sigma_-(z)$ equals \be \Sigma_-(z)
=\left(H_{\rm else}' + \frac{\Delta}{2(z-\Delta)}\left(-A +
B\right)^2\right) \otimes \ketbra{0}_w +
O\left(\frac{||V||^3}{(z-\Delta)^2}\right). \ee
We can expand the self-energy around $z=0$ and identify $H_{\rm eff}=H_{\rm target}\otimes \ket{0}\bra{0}$. This gives
\be
||\Sigma_-(z)-H_{\rm eff}||=O\left(\frac{|z| r^2}{\Delta^2}\right)+O\left(\frac{||V||^3}{\Delta^2}\right)+O\left(\frac{|z| ||V||^3}{\Delta^3}\right).
\ee
In order for Theorem \ref{kkrtheorem} to apply the following must hold: (1) for $z \in
[-\|H_{\rm eff}\|-\eps, \|H_{\rm eff}\| + \epsilon]$, $\Sigma_-(z)$
should be $\epsilon$-close to $H_{\rm eff}$ and (2) $||V|| \leq \Delta/2$. Let us consider how to choose $\Delta$ such that these
conditions are fulfilled.
We can bound $||V|| \leq ||H_{\rm else}'||+\sqrt{2 \Delta}
r \leq \sqrt{\Delta} \left(||H_{\rm else}'||+\sqrt{2} r\right)$.
We will choose $\Delta$ such that $|z| < \Delta$. Then, using the bound on $||V||$ gives
\be
||\Sigma_-(z)-H_{\rm eff}|| \leq O\left(\frac{r^2}{\Delta}\right)+O\left(\frac{(||H_{\rm else}'||+\sqrt{2} r)^3}{\Delta^{1/2}}\right).
\label{eq:sigmabound}
\ee
Let us choose \be \Delta=\left(||H_{\rm else}'||+C_2
r\right)^6/\epsilon^2, \label{choicedelta} \ee for some constant $C_2 \geq \sqrt{2}$.
This choice lets us bound the last term in Eq.~(\ref{eq:sigmabound}) by $O(\epsilon)$.
Since $\Delta^{-1} \leq \frac{\epsilon^2}{C_2 r^6}$, we can bound the first term in Eq.~(\ref{eq:sigmabound}) by $O(\epsilon^2)$. Let us verify the second condition $||V|| \leq \Delta/2$ with this choice of $\Delta$.
We have indeed
\be
\frac{||V||}{\Delta} \leq \frac{\epsilon}{(||H_{\rm else}'||+\sqrt{2} r)^2} \leq \epsilon.
\ee
Consider the conditions on $|z|$, i.e. $z \in
[-\|H_{\rm eff}\|-\eps, \|H_{\rm eff}\| + \epsilon]$ and $|z| < \Delta$. Since
$||H_{\rm eff}||\leq ||H_{\rm else}'||+2 r^2$, we can
consider the interval $|z| \leq ||H_{\rm else}'||+2 r^2 +\epsilon$. For sufficiently small
$\epsilon$ we have (using $\max(||H_{\rm else}'||,r) \geq 1$)
\be
\frac{|z|}{\Delta}=\frac{\epsilon^2 (||H_{\rm else}'||+2 r^2+\epsilon)}{(||H_{\rm else}'||+C_2 r)^6} \leq
O(\epsilon^2) < 1.
\ee
Thus for the choice of $\Delta$ as in Eq.~(\ref{eq:sigmabound}) we have
$\Sigma_-(z) = H_{\rm target} \otimes \ketbra{0}_w +O(\epsilon)$.
From Theorem \ref{kkrtheorem} it follows that $ |\lambda(H_{\rm
eff})-\lambda(\tilde{H})|=O(\epsilon)$.
When $||H_{\rm else}'||$, $r$ and $1/\eps$ are polynomial
in $n$ ($n$ is the number of qubits of $H_{\rm target}$), it is clear from Eq.~(\ref{choicedelta}), that the norm of
the gadget Hamiltonian $\tilde{H}$ which uses $\Delta$ is
polynomially larger than the norm of the effective Hamiltonian. This
implies that the gadget can only be used a {\em constant} number of
times {\em in series} if norms have to remain polynomial.

We will use this type of gadget in parallel, that is, in many places in an interaction graph
at once. Let us explain how this happens in detail and argue that the local gadgets operate
independently, i.e. there are no cross-gadget contributions to 2nd order in the perturbation.
Let $H_{\rm target}=H_{\rm else}-\sum_{i=1}^k H_{\rm target}^i$ where $H_{\rm target}^i=(-A_i+B_i)^2/2$ for some operators $A_i$ and $B_i$. $H_{\rm else}$ contains all
interactions that are not generated perturbatively in addition to the compensating terms
$A_i^2/2$ etc., similar as above. We introduce $k$ mediator qubits $w_1 \ldots w_k$ and choose
$\tilde{H}=\sum_i H_i+V$ where $H_i=\Delta \ketbra{1}_{w_i}$ and
$V=H_{\rm else}+\sqrt{\Delta/2} \sum_i (-A_i+B_i)\otimes X_{w_i}$.

The degenerate ground-space ${\cal L}_-$ of $H$ has all mediator
qubits $w_1 \ldots w_k$ in the state $\ket{0}$. Let $h(x)$ be the
Hamming weight of a bit-string $x \in \{0,1\}^k$ of the qubits $w_1
\ldots w_k$. We have the following: $G_+=\sum_{x \neq 00\ldots
0}\frac{\ketbra{x}}{z-h(x)\Delta}$, $V_{--}=H_{\rm else}\otimes
\ketbra{00 \ldots 0}$ and \be V_{+-}=\sqrt{\Delta/2} \sum_{i} (-A_i
+ B_i)\ket{00 \ldots 1_i \ldots 0}\bra{00 \ldots 0}, \ee where
$\ket{00 \ldots 1_i \ldots 0}$ has qubit $w_i$ in the state
$\ket{1}$. To second order in the perturbation $V$, there are no
cross-gadget terms in $\Sigma_-(z)$. Thus the self-energy
$\Sigma_-(z)$ to second order equals \be \Sigma_-(z)= \left(H_{\rm
else}+\frac{\Delta}{2(z-\Delta)}\sum_i\left(-A_i +
B_i\right)^2\right)\otimes \ketbra{00 \ldots
0}+O\left(\frac{||V||^3}{(z-\Delta)^2}\right). \ee Choosing
$\Delta={\rm poly}(n)/\epsilon^2$ for some sufficiently large ${\rm
poly}(n)$ gives \be \Sigma_-(z) = H_{\rm target} \otimes \ketbra{00
\ldots 0} +O(\epsilon). \ee

We need to use the parallel application of this gadget twice in order to reduce
the ground-state energy problem of our 5-local Hamiltonian to that of a 3-local Hamiltonian; one application
results in a 4-local Hamiltonian, another one reduces it to 3. Similarly, any $k$-local Hamiltonian
for constant $k$ can be reduced to a 3-local Hamiltonian by these means.
A 3-to-2-local reduction can be carried out using the gadget in \cite{KKR:hamsiam}. However an
alternative construction exists which we now explain.

\noindent {\sc 3-to-2-local gadget}. Assume that we have a target Hamiltonian $H_{\rm target}=A \otimes B\otimes C+H_{\rm else}$.
The idea is to generate the 3-local term $A \otimes B \otimes C$ by using
perturbative effects up to third order. As before one introduces
a mediator qubit $w$ whose ground-state is $\ket{0}$ for the unperturbed operator. And, as before,
we have perturbations proportional to $A \otimes X_w$ and $B \otimes X_w$ which can flip the mediator
qubit. We also have a perturbation $V$ which contains a term proportional to $C \otimes \ketbra{1}_w$ which implies that there is an interaction
with $C$ {\em if} the mediator qubit is `excited'. Thus, the second-order perturbative corrections give
us terms proportional to $A \otimes B$ whereas third-order corrections gives us the desired
$A \otimes B \otimes C$ (and some additional 2-local terms). More precisely,
let $H_{\rm target}=H_{\rm else}+A \otimes B \otimes C$. Let $r=\max( ||A||, ||B||, ||C||)$.
We choose $H=\Delta \ketbra{1}_w$ and
\be
V=H_{\rm else}+V_{\rm extra}-\Delta^{2/3} C \otimes \ketbra{1}_w+\Delta^{2/3}(-A+B)\otimes X_w/\sqrt{2}
\ee
where the additional 2-local compensating term is $V_{\rm extra}=\Delta^{1/3}(-A+B)^2/2+(A^2+B^2) \otimes C/2$.
One can show that
\be
\Sigma_-(z)=[H_{\rm else}+A \otimes B\otimes C] \otimes \ketbra{0}_w+O(|z|\Delta^{-2/3})+O(\Delta^{-1/3}).
\ee
For sufficiently large $\Delta$ and $|z| \leq ||H_{\rm else}||+O(r^3)+\epsilon$ we make $\Sigma_-(z)$
sufficiently close to $H_{\rm target}\otimes \ketbra{0}$.

The important conclusion of this section is that one can derive a 2-local Hamiltonian on a spatially
sparse graph for which the ground-state energy problem is QMA-complete. The interaction graph is
restricted because the perturbation gadgets preserve the spatial restrictions of the
original hypergraph of the 5-local Hamiltonian.

\subsection{More Mediator Qubit Gadgetry}
\label{compgadgets}

For our next round of reductions we need to describe
some different uses of the subdivision gadget acting on 2-local interactions. In the following we will assume that every edge in the
interaction graph is a Pauli edge. It may thus be that the interaction graph contains other edges between
the same vertices, each edge associated with a different product of Paulis. The Pauli degree of a vertex
is then the number of Pauli edges that are incident on this vertex.

\begin{figure}[h]\begin{center}\includegraphics[scale=.30]{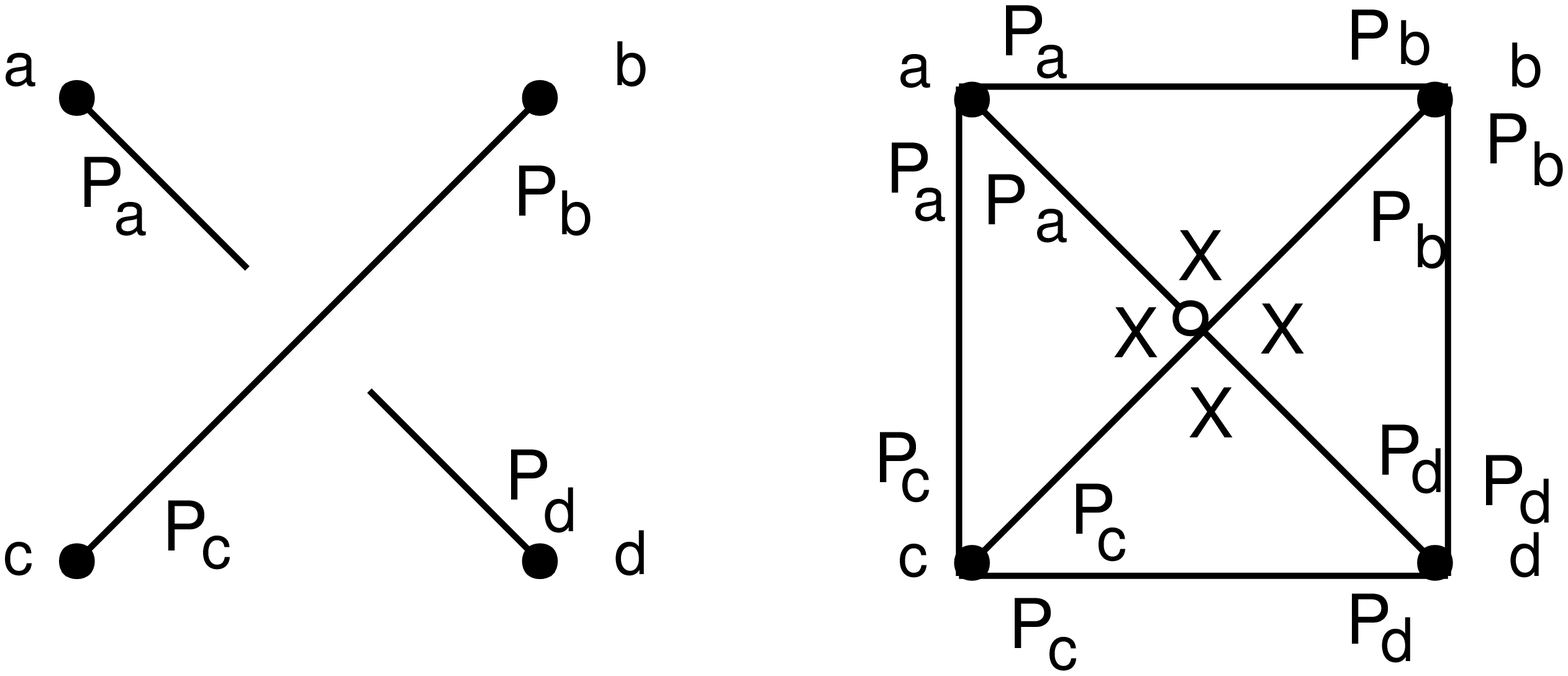}\end{center}\fcaption{\label{fig:cross}
Cross gadget. A crossing between two edges is removed by placing a
mediator qubit in the middle. Additional edges $ab$, $ac$, $bd$ and
$cd$ are created. }\end{figure}

{\sc The Cross Gadget.} For the Cross Gadget we assume that we
have a graph $G$ which, when embedded in the plane, contains two
crossing edges such as in Fig. \ref{fig:cross}. Assume that the
operator on edge $ad$ is
 $\alpha_{ad} P_a \otimes P_d$ and on edge $bc$ we have $\alpha_{bc} P_b \otimes P_c$.
Our desired Hamiltonian is \be H_{\rm target}=H_{\rm
else}-(-\alpha_{ad} P_a-\alpha_{bc} P_b+P_c+P_d)^2/2. \ee It is
clear that the last term in this Hamiltonian generates the desired
crossing edges $\alpha_{ad} P_a \otimes P_d$ and $\alpha_{bc} P_b
\otimes P_c$ in addition to other operators on the edges $ab$, $bd$,
$cd$ and $ac$. Thus $H_{\rm else}$ is a sum of all other operators
associated with the original graph $G$ and a set of operators on the
edges around the cross, see Figure \ref{fig:cross}, which are meant
to cancel the extra operators generated by the last term in $H_{\rm
target}$. As before we set $\tilde{H}=H+V$ with \be H=\Delta
\ketbra{1}_w, \;\; V =H_{\rm else}+ \sqrt{\Delta/2}
\left(-\alpha_{ad}P_a-\alpha_{bc} P_b+P_c+P_d\right) \otimes X_w,\ee
and the analysis follows as for the subdivision gadget. Note that if
there are no edges $ab$, $bd$, $cd$, or $ac$ in $H_{\rm target}$,
there will be such edges in $\tilde{H}$, as indicated in Fig.
\ref{fig:cross}.

\begin{figure}[h]\begin{center}\includegraphics[scale=.30]{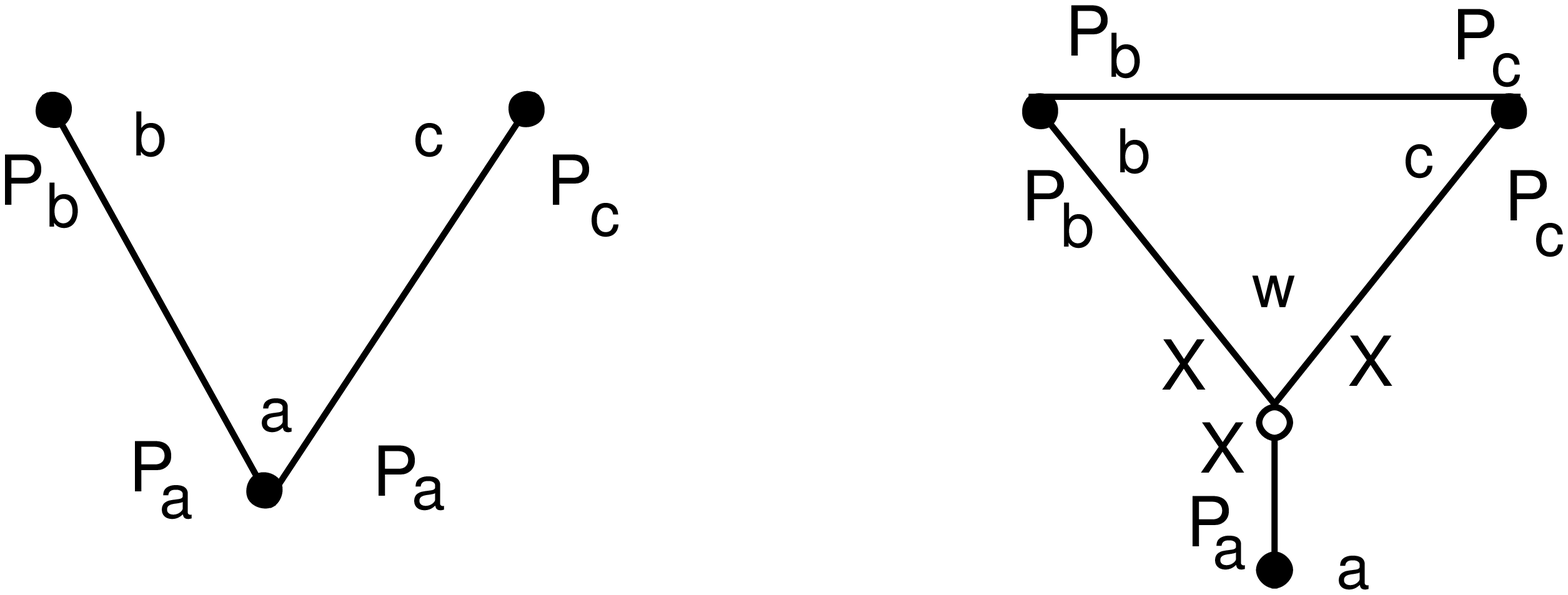}\end{center}\fcaption{\label{fig:y}
Fork gadget. Two edges of the same type at vertex $a$ are merged by
the placement of a mediator qubit $w$. The additional edge $bc$ is
created. }\end{figure}

{\sc The Fork Gadget.} For the Fork gadget we have a subgraph as in
Fig. \ref{fig:y} where the operator on edge $ab$ is $\alpha_{ab} P_a
\otimes P_b$ and on edge $ac$ it is $\alpha_{ac} P_a \otimes P_c$.
The Fork gadget merges the 2 edges coming from vertex $a$ at the
cost of creating an additional edge between $b$ and $c$. Our desired
Hamiltonian is \be H_{\rm target}=H_{\rm else}-(P_a-\alpha_{ab}
P_b-\alpha_{ac}P_c)^2/2, \ee where $H_{\rm else}$ contains all other
terms not involving edge $ab$ and $ac$. We take \be H=\Delta
\ketbra{1}_w, \;\; V =H_{\rm else}+ \sqrt{\Delta/2}
\left(P_a-\alpha_{ab} P_b-\alpha_{ac}P_c\right) \otimes X_w, \ee and
the analysis follows as before.

\begin{figure}[h]\begin{center}\includegraphics[scale=.30]{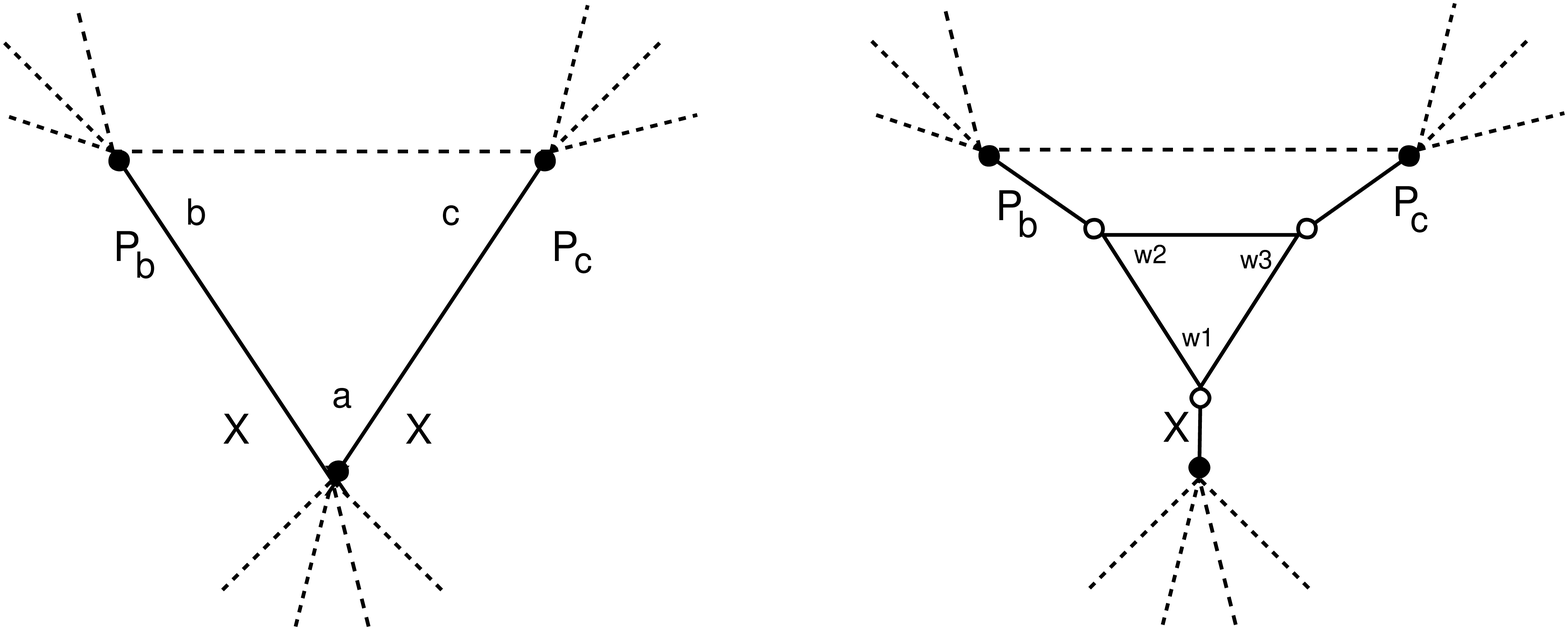}\end{center}\fcaption{\label{fig:triangle}
Triangle Gadget. We first subdivide edges $ab$ and $ac$ and then
apply the Fork gadget on vertex $a$. This give rise to a `mediator
triangle' such that vertices b and c have the same degree as before
and vertex a has reduced its degree by 1.}\end{figure}

\indent {\sc The Triangle Gadget} The Fork
gadget can also be used in order to reduce the degree of a vertex, see
Fig. \ref{fig:triangle}; this is achieved by applying the Fork gadget together with the
subdivision gadget {\em in series}. We first apply a
subdivision gadget on the edges $ab$ and $ac$. Then we apply the
Fork gadget on vertex $a$, thus generating the inner triangle in
Fig. \ref{fig:triangle}.

\section{{\small 2-LOCAL HAMILTONIAN} on a 2-D Square Lattice}
\label{sec:square}

With these tools in place, we are ready to state the reduction which we obtain by
applying the gadgets in the previous section. Together with our previously argued
5-local to 2-local reduction, this Lemma implies Theorem \ref{theo:degree4}.

\begin{figure}[h]\begin{center}\includegraphics[scale=.30]{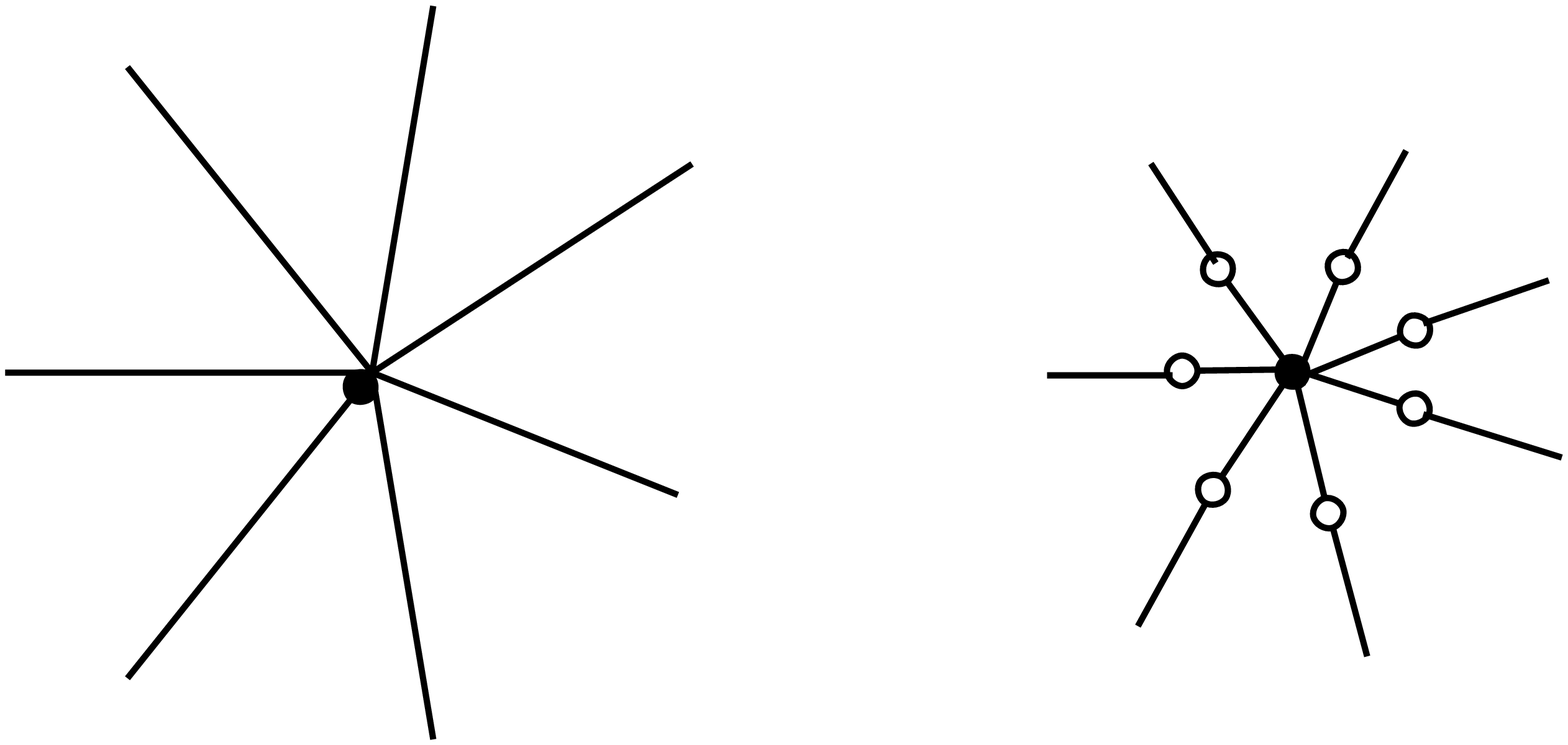}\end{center}
\fcaption{\label{fig:localvertex}Localizing a vertex.}\end{figure}

\begin{figure}[h]\begin{center}\includegraphics[scale=.30]{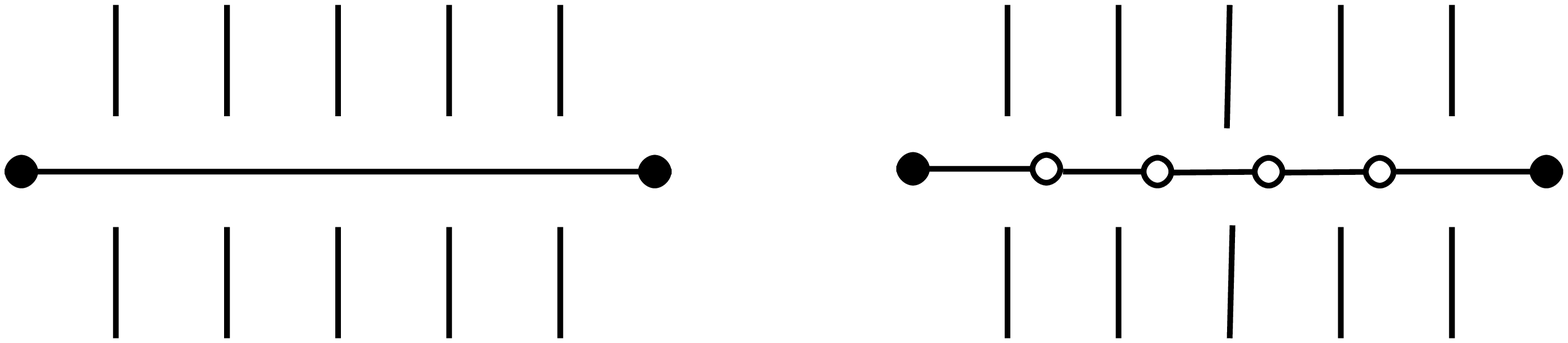}\end{center}\fcaption{\label{fig:logsub}
An edge that crosses $C$ other edges is subdivided $\lceil \log C
\rceil$ times by inserting a mediator qubit.}\end{figure}

\begin{figure}[h]\begin{center}\includegraphics[scale=.30]{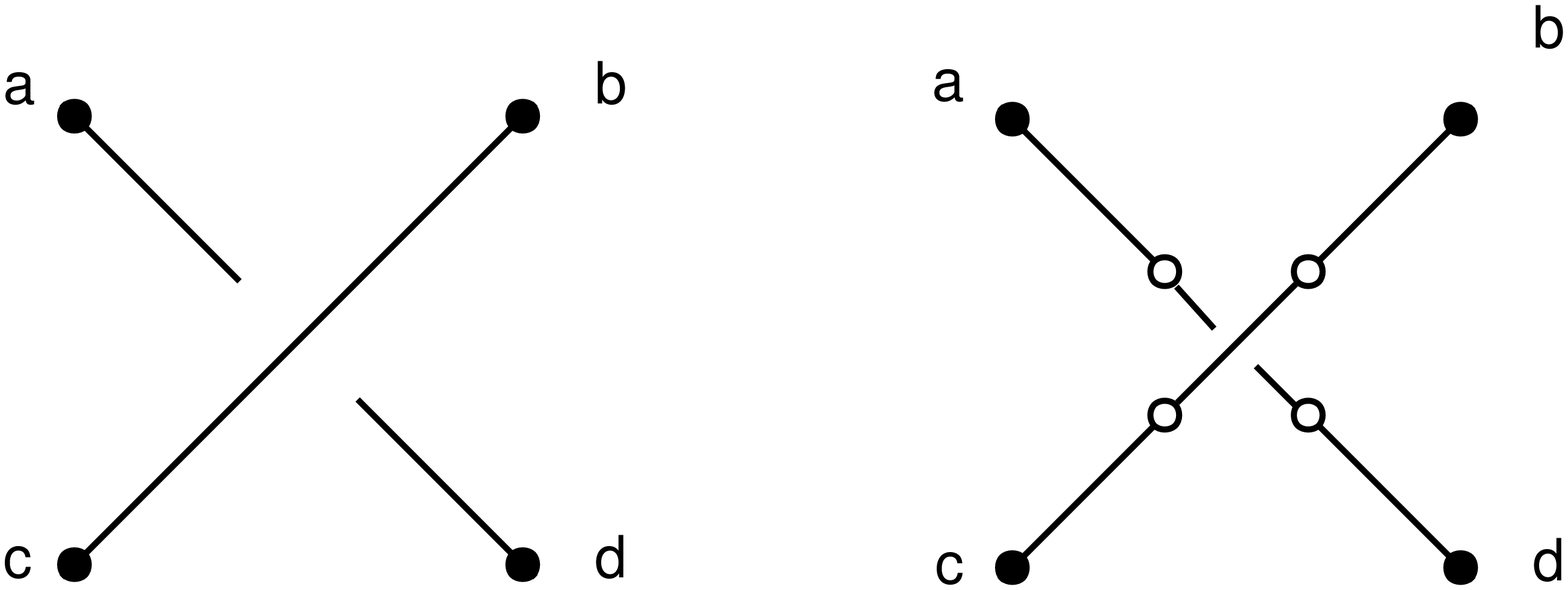}\end{center}
\fcaption{\label{fig:localcross2} Localizing a crossing by applying the
subdivision gadget four times.}\end{figure}

\begin{lemma}\label{lem:planar}Let $H_{G}$ be a 2-local Hamiltonian related to a
spatially sparse graph $G=(V,E)$ with $|V|=n$ and where $H_{G}=\sum_{e \in E}
H_e+\sum_{v\in V}H_v$ such that $\|H_e\| \leq {\rm poly}(n)$ and
$\|H_v\| \leq {\rm poly}(n)$. For any $\epsilon > 0$ there exists a
graph $G^{\rm sim}$ which is planar with maximum Pauli degree at
most 3 and a polynomially bounded 2-local Hamiltonian $H_{G^{\rm
sim}}$ such that \be |\lambda(H_G)-\lambda(H_{G^{\rm
sim}})|=O(\epsilon). \label{redg} \ee Moreover, there is a planar
straight-line drawing of $G^{\rm sim}$ such that all edges in
$G^{\rm sim}$ have length $O(1)$, and all angles between adjacent
edges are $\Omega(1)$.\end{lemma}
\proof{
\begin{itemize}
\item We use the subdivision gadget in order to localize each
vertex with Pauli degree more than 3, see Fig.
\ref{fig:localvertex}.
Then we are ready to reduce the Pauli degree (which is some constant) of these vertices.
\item Consider the set of vertices with Pauli degree more than 3. We are going to apply
the Triangle gadget to all these high degree vertices in the
following way. We first apply the subdivision gadget to all edges
that we intend to merge using the Fork gadget; we can do this in one
parallel application. We do this so that the triangle gadgets that
we will apply in parallel never act on the same edges. Then, for a
vertex with $X$-degree $d_x$, $Y$-degree $d_y$, $Z$-degree $d_z$ we
do the following. We pair the $X$-edges and apply to each pairing a
Fork gadget. This means we have reduced the $X$-degree to $\lceil
d_x/2 \rceil$. In parallel we pair the $Y$-edges and the $Z$-edges
using the Fork gadget, halving their degrees. We do this single
perturbative step in parallel for all high-degree vertices in the
graph. We repeat this Triangle gadget process $O(1)$ number of times
(since the maximum degree initially was $O(1)$) until the total
Pauli degree of every vertex is at most $3$. Since the initial
degree of every vertex was $O(1)$, the number of additional
crossings that we generate per edge is constant.
\item Next, we reduce the number of crossings per edge, by subdividing
each edge a constant number of times, see Fig. \ref{fig:logsub}. Every subdivision is
done in parallel on all
edges of the graph that need subdividing.
\item Then we use the subdivision gadget to {\em localize} each
crossing, see Fig. \ref{fig:localcross2}. We apply the subdivision
gadget in parallel on every crossing in the graph and we repeat the
process $4$ times so that for all crossing
edges $ab$, $cd$, the quadrilateral $acbd$ contains only these points and the crossing edges.
\item We apply the Cross gadget, see Fig. \ref{fig:cross}, in parallel
to every localized crossing in order to remove the crossing. Note that due to the localization
step the cross-gadget only involves mediator qubit vertices with degree at most 2.
Thus the cross-gadget generates additional 2-local terms around the square, but the total
Pauli degree of the resulting vertices is at most 4. Note that these vertices with degree 4 are all mediator
qubits which only have non-zero X-degree (and zero Y- and Z-degree).
\item On all mediator qubits with X-degree 4 we apply the Triangle gadget reducing the degree to 3.
Since the triangle gadget generates mediator qubits with X-degree 3 we cannot do any further reductions.
\end{itemize}
Thus in this final Hamiltonian there are no vertices with Pauli degree more than 3 and the graph is planar.
Theorem \ref{kkrtheorem} is used in every gadget application to give the final result, Eq.~(\ref{redg}).
Note that by this reduction all original system qubits have Pauli degree at most 3 by having X-degree, Y-degree and
Z-degree ranging from 0 to 1. The mediator qubits have X-degree ranging from 2 to 3 and 0 Y- and Z-degree.}

\subsection{Representation on a 2-D Square Lattice}

\begin{figure}[h]\begin{center}\includegraphics[scale=.20]{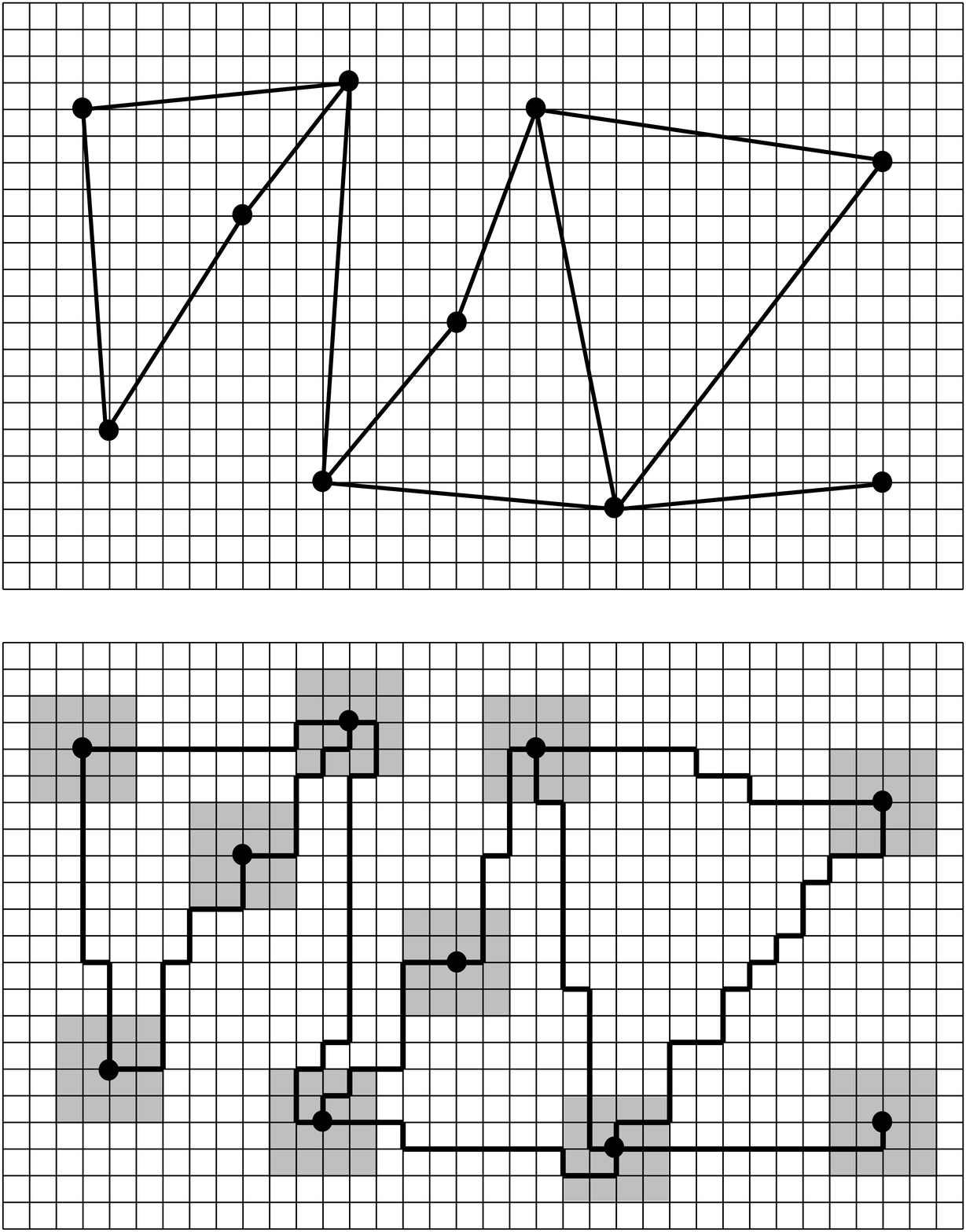}\end{center}
\fcaption{\label{fig:lattice}A planar graph of maximal degree $\leq
3$ and its representation in the lattice. In the gray squares, the
paths are rerouted to avoid crossings.}\end{figure}

Any planar graph $G=(V,E)$ with maximal degree $3$ in which the
(straight-line) edges have length $O(1)$ and adjacent edges form
an angle of $\Omega(1)$ can be {\em represented} on a planar
square lattice in the following sense: each vertex $a$ of $G$ is
mapped to some lattice site $\phi(a)$ inside the square
$[-O(|V|),O(|V|)]^2$, and each edge $ab$ of $G$ is mapped to a
lattice path $\phi(ab)$ of length $O(1)$ from $\phi(a)$ to
$\phi(b)$ that does not cross any other vertices or any other
path. To see this, one can look at Fig. \ref{fig:lattice} or
follow these steps: draw a fine square grid on the plane. If the
spacing between points on the grid is small enough, moving each
vertex $a$ of $G$ to a vertex in the lattice (and redrawing the
edges) still leaves the graph planar, with $O(1)$-length edges and
$\Omega(1)$ angles. Now for each edge, draw a lattice path that
stays close to the edge. If the grid is fine enough, these paths
can never cross outside an $O(1)$-size square (indicated in grey
in Fig. \ref{fig:lattice}) around the vertices of the graph,
because of the angle condition. By further refining the grid if
necessary, one can reroute each of the paths stemming out of a
vertex $a$ inside of $a$'s square, so that no two different paths
collide. It is easy to see that we only need the grid to have
spacing $\Omega(1)$, and that all the other conditions above are
satisfied.

Clearly, this embedding can be found efficiently, given the
adequate embedding of $G$. If $H$ is a Hamiltonian that has $G$ as
(Pauli) interaction graph, one can use the subdivision gadget
$O(1)$ times in parallel to map each edge $ab$ to a path of the
same length as $\phi(ab)$. The Hamiltonian $\tilde{H}$ thus
obtained has interaction graph $\phi(G)$ and $\lambda(\tilde{H})$
is $O(\epsilon)$-close to $\lambda(H)$. These arguments together with our previous results
and Lemma \ref{lem:planar} prove Theorem \ref{theo:squarelat}.

\section{Universal Quantum Adiabatic Computation}\label{sec:adiabatic}

In Ref.~\cite{KKR:hamsiam} the authors show that their
perturbation-theoretic reduction of {\small 3-LOCAL HAMILTONIAN} to
{\small 2-LOCAL HAMILTONIAN} also reduces $3$-local adiabatic
computations to $2$-local ones. The goal of this Section is to show
that an analogous result can be carried out in the present context,
namely that $2$-local Hamiltonians with nearest-neighbor
interactions on qubits on a 2D lattice suffice for universal
adiabatic quantum computation.

Let us describe in more detail what our goal is. We will construct a
(classically) poly-time computable map $\Phi$ that takes as input a
classical description $\langle Q\rangle$ of a quantum circuit $Q$
and outputs a description of an adiabatic quantum computation on a
2D lattice. Suppose $Q$ acts on $n$ qubits and has $T$ gates. Then
$$\Phi(\langle Q\rangle)=(\langle
H_0\rangle, \dots, \langle H_p\rangle).$$ Here
\begin{enumerate}
\item $p\in\mathbb{N}$ is a constant independent of $Q$;
\item $N$, the number of qubits on which $H_i$ acts is ${\rm poly}(n,T)$;
\item for each $i\in\{0,\dots,p\}$, $\langle H_i\rangle$ describes a
$2$-local nearest-neighbor Hamiltonian on qubits, acting on the same
subset of $N={\rm poly}(n,T)$ sites of the square lattice;
\item$\|H_i\|={\rm poly}(n,T)$ for all
$i\in\{0,\dots,p\}$;
\item Let $$H(s)=\sum_{i=0}^p s^i\,H_i\, (s\in[0,1]).$$ The spectral gap between the
ground-state and first excited state of $H(s)$ is $1/{\rm
poly}(n,T)$ for all $s$.
\item The ground-state of $H(0)$ is $\ket{0}^{\otimes
N}$ and the ground-state of $H(1)$ encodes the result of the
computation of $Q$ on input $\ket{0}^{\otimes n}$ (a more precise
description is given in \cite[Section 7]{KKR:hamsiam} or
\cite{ADLLKR:adia}).\end{enumerate}

Of course, all occurrences of ${\rm poly}$ above correspond to fixed
polynomials that do not depend on $Q$. Notice that for any
Hamiltonian satisfying the above conditions one has that
$$\sup_{s\in[0,1]}\left\|\frac{d^jH(s)}{ds^j}\right\|\leq {\rm poly}(n,T), \,\mbox{for all }j=0,1,\ldots $$
This is sufficient to ensure that adiabatic computation implemented
by $H(s)$, starting from $\ket{0}^{\otimes N}$, appropriately
simulates the quantum circuit $Q$, that is, in polynomial time \cite{AR:adia}.
Note that the usual adiabatic computation, e.g. the universal adiabatic computation in \cite{ADLLKR:adia}, has
$p=1$.


There are several ways to map a circuit $Q$ to a corresponding
$H(s)$. One could modify the 5-local construction in this paper in
order to show that one can do universal quantum computation using a
quantum adiabatic computation with a 5-local Hamiltonian on a
spatially sparse graph. Then we could apply the perturbation gadgets
to derive a 2-local Hamiltonian with similar properties. However, an
easier route to the desired result is the following. In
\cite{ADLLKR:adia} it was shown how to map a circuit $Q$ to a
corresponding Hamiltonian $H^{(6)}(s)$ on a 2D lattice. That
construction satisfies all but one of the above requirements, as it
acts on $6$-dimensional qudits rather than qubits. However, we can
embed 6-dimensional qudits in states of 3 qubits. This
implies that the 2-local interactions between these particles will
be mapped onto 6-local interactions. Then we can apply the
perturbation gadgets to `massage' this Hamiltonian on a spatially
sparse hypergraph to a 2-local Hamiltonian as we have done in our QMA construction.\\

\indent Let us first review the 6-dim particle Hamiltonian, see Sec.~4.2 in \cite{ADLLKR:adia}. The four phases of the particles, the
unborn, the first, second and dead phase, can be described by two
qubits in the states $\ket{\rm unborn}=\ket{00}$,$\ket{\rm
first}=\ket{01}$, $\ket{\rm second}=\ket{10}$ and $\ket{\rm
dead}=\ket{11}$. The third qubit holds the actual computational
degree of freedom. In the 6-dimensional representation of the unborn
and dead phase the computational degree of freedom is assumed to be
fixed. If we represent those states as 3-qubit states, we fix the
third qubit to be in the state $\ket{0}$. Hence we obtain 6 states:
2 `first' states $\ket{010},\ket{011}$, two `second' states
$\ket{100},\ket{101}$ and one unborn state $\ket{000}$ and one dead
state $\ket{110}$. With this mapping the entire Hamiltonian in Sec.~4.2 in \cite{ADLLKR:adia} can be rewritten in terms of $6$-local
interaction between qubits. Since we embed the 6-dim particle
Hamiltonian in a higher dimensional space, we need to make sure that
states outside the embedded space (i.e. $\ket{001}$ and $\ket{111}$)
are penalized in the Hamiltonian, i.e. do not contribute to the
ground-space. In Table 1 in \cite{ADLLKR:adia} a list of forbidden
configurations is given. In this list we can replace every unborn
state by two unborn states $\ket{{\rm unborn},1}$ and $\ket{{\rm
unborn},0}$ and similarly for the dead states. This implies a small
modification of $H_{\rm clock}''$. As a consequence we get that the
space of legal shapes ${\cal S}$ is the same for this embedded
Hamiltonian as for the original 6-dim particle Hamiltonian. It then
follows that one can apply Lemma 4.6 and 4.7 bounding the spectral
gap of the Hamiltonian in the space of legal states. We note in
passing the Hamiltonian $H^{(6)}(s)$ has only linear terms in $s$ (that
correspond to $p=1$ above) and terms independent of $s$ (corresponding to
$p=0$).

Our second step is to analyze how this desired 6-local Hamiltonian
can be implemented using a 2-local Hamiltonian on a 2D lattice. It
is clear that one can apply the perturbation gadgets in Section 3
and 4 of this paper and map a 6-local Hamiltonian on a spatially
sparse hypergraph onto a 2-local Hamiltonian on a subgraph of the 2D
lattice. To go from a 3-local to a 2-local Hamiltonian we will use
our alternative 3-to-2-local gadget described in Section \ref{sec:medqubits}. One needs to show the following properties of
the perturbation method in order for these reductions to work:
\begin{enumerate}
\item The 2-local adiabatic path Hamiltonian $H^{(2)}(s)$ obtained
through the perturbation gadgets simulates the 6-local adiabatic
path Hamiltonian. This implies that the ground-state of the 2-local
Hamiltonian should be approximately the ground-state of the desired
6-local Hamiltonian and the gap for the 2-local Hamiltonian is
approximately the gap of the 6-local Hamiltonian. This requires
showing that the perturbative method that we employ does not only
reproduce the lowest-eigenvalue but also the ground-state and the
gap above the ground-state.
\item One needs to verify that $H^{(2)}(s)$ is of
the form $\sum_{i=0}^p s^i\, H_i$, with $p$ constant and
$\max_i\|H_i\|\leq {\rm poly}(n,T)$.
\end{enumerate}

Our 2-local simulator Hamiltonian $H^{(2)}(s)$ is determined by
applying the perturbative gadgets in Sections \ref{sec:medqubits} and
\ref{compgadgets}, on the 6-local Hamiltonian $H^{(6)}(s)$. In \cite{KKR:hamsiam} it was shown how to generate,
not only the lowest eigenvalues, but also the ground-state with the perturbative technique.
This implies that both the ground-state of the target
Hamiltonian as well as the gap above this ground-state can be
generated perturbatively. Since the total number of applications of
the perturbation theory is constant, one can apply this argument for
each step and thus show that the 6-local target Hamiltonian can be
effectively generated by a simulator Hamiltonian $H^{(2)}$.

We now fulfill our second task, i.e. we show that
$H^{(2)}(s)=\sum_{i=0}^p s^i\, H_i$, with $p$ constant and $\|H_i\|$
polynomially bounded. In \cite{KKR:hamsiam} such
arguments were developed for the 3-to-2 local perturbation gadget
and basically identical arguments can be given here. The original Hamiltonian $H^{(6)}$ is at most
linear in $s$. If a gadget is applied on a term which is linear in $s$, for example
a 6-local term such $s A \otimes B=A(s)\otimes B$, we obtain a new Hamiltonian of which the terms are
at most quadratic in $s$. Similarly each application of the perturbation gadgets takes a
Hamiltonian $H'(s)=\sum_{j=0}^{p'} s^j\,H'_j$ to another Hamiltonian
$H''(s)=\sum_{i=0}^{p''} s^i\,H''_i$ where $p''\leq 2p'$. Assuming
that the norm of each $H'_i$ is polynomial in $n$ and $T$, then the norms
of each $H''_j$ are also ${\rm poly}(n,T)$. Thus the final
Hamiltonian $H^{(2)}$, obtained after a constant number of gadget
applications, is indeed of the desired form.

\section{Discussion and Acknowledgements}
The drawback of the reductions performed by our perturbation theory
method is that the 2-local Hamiltonian that we construct has large
variability in the norms of the 2-local terms. In other words, 2-local terms
have constant norm whereas others can be fairly high degree
polynomials in $n$.  Such dependence on $n$ may be undesirable from
a practical point of view, e.g. if one wants to perform universal
adiabatic quantum computation.

It is possible that a less stringent but still rigorous perturbation theory could be developed
in which only the expectation values of {\em local} observables with respect to
the ground-space are perturbatively generated. If such expectation values are reproduced
with constant accuracy (not scaling as $1/{\rm poly}(n)$), then the perturbation theory
need not be accurately reproduce the entire ground-space as in Lemma 3.
For adiabatic quantum computation this method would suffice since one can measure
a single output qubit to extract the answer of the computation.

One of the reasons why finding QMA-complete problems is of interest
is that it may give us a hint at what problems can be solved in BQP.
One example is the unresolved status of the 2-local Hamiltonian
problem on qubits in one dimension. Another
example is to find a quantum extension of classical 2-local
Hamiltonian problems which can be solved efficiently. We thank
David DiVincenzo for an inspiring discussion about superexchange. We would like to thank Sergey Bravyi for pointing
out an improvement in the proof of Lemma 2. We
acknowledge support by the NSA and the ARDA through ARO contract
number W911NF-04-C-0098.

\appendix

\section{General Perturbation Theorem}

In order to give a more complete background in the perturbation
method we will prove in Theorem \ref{thm:perturbation} that under the right conditions the entire
operator $\tilde{H}|_{< \lambda_*}$ is approximated by $H_{\rm
eff}$, not only its eigenvalues. In
Ref.~\cite{KKR:hamsiam} a similar result was proven, namely that the
ground-state of $\tilde{H}$ is approximately the ground-state of
$H_{\rm eff}$. We extend their result to the case when the
ground-space is degenerate in Lemma \ref{lem:groundpert} of this
Appendix. To a certain extent our proof-technique is similar to the
one used in Ref.~\cite{KKR:hamsiam}, however we will use complex $z$
and contour integration in parts of the proofs.

Some of our notation has been given in Section \ref{sec:gadgets} for
the specific cases considered in this paper. Here we consider the
more general setting as defined in Ref.~\cite{KKR:hamsiam}.

Assume that $H$ and $V$ are operators acting on the Hilbert space
${\cal L}$ and $\tilde{H}=H+V$. $H$ has a spectral gap $\Delta$ such that no eigenvalues
lie in the interval $[\lambda_*-\Delta/2,\lambda_*+\Delta/2]$ for some cutoff $\lambda_*$. Let
${\cal L}_-$ (resp. ${\cal L}_+$) be the span of all eigenvectors of $H$ whose eigenvalues are less than $\lambda_*$ (respectively larger than $\lambda_*$). We will use the
{\em resolvent} $G(z)\equiv (zI-H)^{-1}$ of $H$ with {\em complex} $z\in\bbC$ and
 let $\tilde G(z) = (zI-\tilde H)^{-1}$ be the resolvent of $\tilde H$. The definition of the self-energy $\Sigma_-(z)$
 is given by
 \be
\Sigma_-(z)=zI_--\tilde{G}_{--}^{-1}(z).
\label{eq:defs}
 \ee
see also Eqs. (\ref{selfe})-(\ref{pertexpand}).

The perturbation theory result of Kempe {\em et al.} states that under
suitable technical conditions, --namely if $\Sigma_-(z)$ is close to a fixed
operator $H_{\rm eff}$ for all $z$ in some range--, all
eigenvalues of $\tilde H = H+V$ that lie below the cutoff
$\lambda_*$ are close to those of $H_{\rm eff}$. Our result shows
that the {\em entire operator} $\tilde H$
restricted to its low-lying energy levels is close to $H_{\rm eff}$ under a slightly stronger
assumption.

\begin{figure}[htb]
\setlength{\unitlength}{0.5cm}
\begin{picture}(30,15)
\thicklines \put(0,7){\line(30,0){30}}

 \put(29,8){$\lambda_*+\Delta/2$} \put(29,7.5){\line(0,-1){1}}
 \put(18,8){$\lambda_*$} \put(18,7.5){\line(0,-1){1}}
\put(14,8){$b$} \put(14,7.5){\line(0,-1){1}}
\put(9,8){$z_0$} \put(9,7.5){\line(0,-1){1}}
\put(4,8){$a$}\put(4,7.5){\line(0,-1){1}}
\put(4,6){\line(30,0){10}}
\put(7,5){${\rm Spec}(H_{\rm eff})$}

\put(9,7){\vector(3,2){5.5}}
\put(11,9.5){$r$}
\put(3,12){$D_r$}
\put(19,5){\vector(-3,1){4.3}}
\put(19.3,5){$b+\epsilon$}

\put(3.5,7.3){\line(0,-1){0.6}}
\put(14.5,7.3){\line(0,-1){0.6}}

\put(9,7){\circle{13}}
\end{picture}
\fcaption{The disk $D_r$ in the complex plane, the spectrum of $H_{\rm eff}$ and
the other parameters in Theorem \ref{eq:thmresult}.}
\label{fig:spec}
\end{figure}

\begin{theorem}\label{thm:perturbation}Given is a Hamiltonian $H$ such that no eigenvalues of $H$ lie between $\lambda_- = \lambda_*-\Delta/2$ and
$\lambda_+ = \lambda_* + \Delta/2$. Let $\tilde{H}=H+V$ where $||V||
\leq \Delta/2$. Let there be an effective Hamiltonian $H_{\rm eff}$
with ${\rm Spec}(H_{\rm eff}) \subseteq [a,b]$, $a < b$. We assume
that $H_{\rm eff}=\Pi_- H_{\rm eff} \Pi_-$. Let $D_r$ be a disk of
radius $r$ in the complex plane centered around
$z_0=\frac{b+a}{2}$. Let $r$ be such that $b+\epsilon < z_0+r <
\lambda_*$ (see Figure \ref{fig:spec}). Let $w_{\rm eff}=\frac{b-a}{2}$.
Assume that for all $z \in D_r$ we have
$\|\Sigma_-(z) -H_{\rm eff}\|\leq \eps.$ Then \be \| \tilde H_{<\lambda_*} - H_{\rm
eff} \| \leq \frac{3 (||H_{\rm eff}||+\epsilon)
\|V\|}{\lambda_+-||H_{\rm eff}||-\eps}+\frac{r(r+z_0)
\epsilon}{(r-w_{\rm eff})(r-w_{\rm eff}-\epsilon)}.
\label{eq:thmresult} \ee
\end{theorem}

Before we prove the theorem, let us make a few comments about how it
can be applied.  We have assumed that $H_{\rm eff}$ has no support
in ${\cal L}_+$; this will be the case in typical applications since
$H_{\rm eff}$ approximates $\Sigma_-(z)$ which has support only on
${\cal L}_-$. It is not hard to modify the theorem if $H_{\rm eff}$
has (necessarily small) support outside ${\cal L}_-$.

The r.h.s in Eq.~(\ref{eq:thmresult}) contains the energy scale $||H_{\rm eff}||$ which is
not invariant under shifts by $\alpha I$. In applying the theorem to a Hamiltonian ${\tilde H}$
one can always shift this Hamiltonian ${\tilde H}$ by $\alpha I$, without changing its eigenvalues or eigenvectors,
such that $H_{\rm eff}$ has a spectrum centered around 0. In that case $||H_{\rm eff}||=
\min_{\alpha} ||H_{\rm eff}+\alpha I||=w_{\rm eff}$, the effective width. Thus one may replace
$||H_{\rm eff}||$ by $w_{\rm eff}$ in the application of the Theorem.

In the construction using mediator qubits, we will choose
$\lambda_-=0$ and thus $\lambda_*=\Delta/2$. ${\cal L}_-$ is the
space in which the mediator qubits are in the state $\ket{00 \ldots
0}$ and $H_{\rm eff}$ is of the form $H_{\rm target} \otimes \ket{00
\ldots 0}\bra{00 \ldots 0}$. In order for the right-hand-side of Eq.
~(\ref{eq:thmresult}) to be small, we need to take the spectral gap
$\Delta$ to be sufficiently large (some ${\rm poly}(n)$). This will
directly bound the first term on the right hand side. Now consider the second term
and the choice for $r$. In our applications
$H_{\rm eff}$ is derived from the perturbative expansion of
$\Sigma_-(z)$. Since $\tilde{H}$ and $H$ on $n$ qubits have norm
${\rm poly}(n)$, the Hamiltonian $H_{\rm eff}$ (related to the
target Hamiltonian) will also have norm ${\rm poly}(n)$. Hence $a,b$
and thus $z_0$ are at most ${\rm poly}(n)$.
Note that we need to take $z_0+r > b+\epsilon$ which implies that
$\Sigma_-(z)$ has to be approximately equal to $H_{\rm eff}$ in a range of $z$
which is larger than what is needed in Theorem \ref{kkrtheorem}.
Secondly, it is necessary that the eigenvalues of $\tilde{H}$ are bounded away from
$\lambda_*$, the difference between the largest eigenvalue below $\lambda_*$ and the smallest
eigenvalue above $\lambda_*$ needs to be at least $1/{\rm poly}(n)$. For our
mediator qubit gadgets, one could take $r$ (for example) to scale as $\Delta^{1/k}$ for some constant
$k > 1$ in order for these conditions to be fulfilled.

\proof{{\bf (of \thmref{perturbation})}We start from Theorem 3 in Ref.~\cite{KKR:hamsiam} (stated as Theorem \ref{thm:perturbation}
in this paper) which shows that
under the assumptions in the Theorem one has $|\lambda_j(\tilde
H_{<\lambda_*} ) - \lambda_j(H_{\rm eff})|\leq \eps$ for each $1\leq
j\leq {\rm dim}({\cal L}_-)$. We can draw a contour $C$ in the
complex plane, the disk $D_r$ in Figure \ref{fig:spec}, that encloses all the eigenvalues of $\tilde H_{<
\lambda_*}$ and none of the higher eigenvalues of $\tilde H$. The radius $r$ needs to be chosen
such that $b+\epsilon < z_0+r$ to include all the eigenvalues of $H_{< \lambda_*}$. At the same time
$z_0+r < \lambda_*$ such that none of the higher eigenvalues of
$\tilde{H}$ are included in the contour integral. Using Cauchy's contour integral formula we can write
\begin{equation}\label{eq:contour}
\tilde H_{<\lambda_*} =\frac{1}{2\pi i}\oint_{C} z \,\tilde{G}(z) \, dz.\end{equation}

The remainder of our proof proceeds in two parts. In the first
part we show that $\tilde H_{<\lambda_*}$ is close to $\Pi_-\tilde
H_{<\lambda_*}\Pi_-$; this is expressed in Eq.~(\ref{eq:finalbound}). In the second part we show
that $\Pi_-\tilde H_{<\lambda_*}\Pi_-$ is close to $H_{\rm eff}$, expressed in Eq.~(\ref{eq:part2}).

{\sc First part.} We
have \bea\|\tilde H_{<\lambda_*} - \Pi_- \tilde
H_{<\lambda_*}\Pi_-\| & = &
\|\Pi_+ \tilde H_{<\lambda_*} \Pi_+ + \Pi_+ \tilde H_{<\lambda_*}\Pi_{-}+\Pi_- \tilde H_{<\lambda_*}\Pi_+\| \nonumber \\
& \leq & 2\,\|\Pi_+ \tilde H_{<\lambda_*}\|+\|\tilde H_{<\lambda_*}
\Pi_+ \|, \eea using standard properties of the operator norm
$||.||$. Let $\tilde \Pi_{<\lambda_*}$ be the projector onto the
space spanned by the eigenvectors of $\tilde H_{<\lambda_*}$ with
eigenvalues below $\lambda_*$. We can insert $\tilde{\Pi}_{< \lambda_*}$ before or after $\tilde H_{< \lambda_*}$
and use that for projectors $P_1,P_2$, $||P_1 P_2||=||P_2 P_1||$, so
that \be ||\tilde H_{<\lambda_*} - \Pi_- \tilde
H_{<\lambda_*}\Pi_-\| \leq 3 \,\|\tilde{H}_{< \lambda_*}\| \, \|
\Pi_+ \tilde{\Pi}_{< \lambda_*}\| \leq 3 \left(||H_{\rm
eff}||+\epsilon\right) \| \Pi_+ \tilde{\Pi}_{< \lambda_*}\|. \ee
In order to bound this, we first derive \be ||\Pi_+ H \tilde{\Pi}_{<
\lambda_*}||=||\Pi_+ H \Pi_+ \tilde{\Pi}_{< \lambda_*}||  \geq
\lambda_+ ||\Pi_+ \tilde{\Pi}_{< \lambda_*}||. \ee On the
other hand, we have \be ||\Pi_+ H \tilde{\Pi}_{< \lambda_*}||  \leq
||\Pi_+ \tilde{H} \tilde{\Pi}_{< \lambda_*}||+||V|| \leq (||H_{\rm
eff}||+\epsilon)||\Pi_+ \tilde{\Pi}_{< \lambda_*}|| +||V||. \ee
Putting the last three equations together gives the final bound
\bea\|\tilde H_{<\lambda_*} - \Pi_- \tilde H_{<\lambda_*}\Pi_-\|
\leq \frac{3 (||H_{\rm eff}||+\epsilon)
\|V\|}{\lambda_+-(||H_{\rm eff}||+\epsilon)}.
\label{eq:finalbound} \eea

{\sc Second part.}
We consider
\begin{equation}\label{eq:contour2}\Pi_-\tilde H_{<\lambda_*}\Pi_- = \frac{1}{2\pi i}\oint_C z\,\Pi_-\tilde G(z)\Pi_- \,
dz,\end{equation}
and recall that $\Pi_-\tilde G(z)\Pi_- =\tilde{G}_{--}(z)=(zI_- -\Sigma_-(z))^{-1}$, Eq.~(\ref{eq:defs}).
By showing that this operator is close to $\Pi_-(zI-H_{\rm eff})^{-1}\Pi_-=(zI_--H_{\rm eff})^{-1}$, we will be able
to deduce that $\Pi_-\tilde H_{<\lambda_*}\Pi_- $ is close to $H_{\rm eff}$.

For all $z \in D_r$, $\|\Sigma_-(z) - H_{\rm eff}\|\leq
\eps$ by assumption. In order to bound
$\|(zI_--\Sigma_-(z))^{-1} - (zI_--H_{\rm eff})^{-1}\|$, we will use the
following \be
||(A-B)^{-1}-A^{-1}||=||(I-A^{-1}B)^{-1}A^{-1}-A^{-1}||\leq
\left((1-||A^{-1}||\,||B||)^{-1}-1\right) ||A^{-1}||,
 \ee
 when $||A^{-1}||\, ||B|| < 1$. We choose $A=zI_--H_{\rm eff}$ and $B=\Sigma_-(z)-H_{\rm eff}$. For $z \in C$ (i.e. on
 the contour) $||A^{-1}||\leq  (r-w_{\rm eff})^{-1}$ and thus $||A^{-1}|| \,||B||
\leq \frac{\epsilon}{r-w_{\rm eff}} \leq 1$. It follows that
$$\sup_{z\in C}\|(zI_--\Sigma_-(z))^{-1} - (zI_--H_{\rm
eff})^{-1}\| \leq \frac{\epsilon}{(r-w_{\rm eff}-\epsilon)(r-w_{\rm
eff})}.$$

Now we will use the following for an operator-valued function $F(z)$
and a contour $C$ with radius $r$ around a real-valued $z_0$: \be
\left\|\frac{1}{2 \pi i}\oint_C \,z\, F(z) \,dz\right\| \leq
r(r+z_0) \sup_{z \in C} ||F(z)||. \ee
Using this bound and the resolvent for $H_{\rm eff}$, we find
that
 \bea ||\Pi_-\tilde H_{<\lambda_*}\Pi_--H_{\rm
eff}|| & = & \left\|\frac{1}{2 \pi i} \oint_C z \left((zI_--\Sigma_-(z))^{-1} - (zI_--H_{\rm
eff})^{-1}\right)\right\| \nonumber \\
& \leq & \frac{r(r+z_0) \epsilon}{(r-w_{\rm eff})(r-w_{\rm
eff}-\epsilon)}.\ \label{eq:part2} \eea where we have used that
$H_{\rm eff}=\Pi_- H_{\rm eff}\Pi_-$. Putting Eqs.
(\ref{eq:finalbound}) and (\ref{eq:part2}) together gives the
desired result, Eq.~(\ref{eq:thmresult}).}

The proof technique used in this theorem can be easily adapted to
prove properties of the low-lying eigenspace of ${\tilde H}$; this is the
content of the following Lemma. For the resulting bound of Eq.~(\ref{eq:groundbound}) to be useful one needs (1) to take $\Delta$
large enough compared to $||V||$ (which bounds the first term in Eq.~(\ref{eq:groundbound}))
and (2) the gap $\Delta_{\rm eff}$ of the effective Hamiltonian $H_{\rm eff}$, defined as $\Delta_{\rm
eff}\equiv \lambda_{1,\rm eff}-\lambda_{0,\rm eff}$, needs to be
bounded away from zero. In particular in the Lemma we can take
$r=\Delta_{\rm eff}-2 \epsilon$ and then the second term in Eq.~(\ref{eq:groundbound}) can be
(upper)-bounded by $\frac{\epsilon \lambda_{1, {\rm eff}}}{(\Delta_{\rm eff}-2 \epsilon)(\Delta_{\rm eff}-3 \epsilon)}$.

If $\Delta_{\rm eff} \geq \frac{1}{{\rm poly}(n)}$ we can thus take a polynomially small
$\epsilon$ to bound the second term in Eq.~(\ref{eq:groundbound}) by some other inverse polynomial.

\begin{lemma}
Given is a Hamiltonian $H$ such that no eigenvalues of $H$ lie between $\lambda_- = \lambda_*-\Delta/2$ and
$\lambda_+ = \lambda_* + \Delta/2$. Let the perturbed Hamiltonian
$\tilde{H}=H+V$ where $V$ is a small perturbation with $||V|| \leq
\Delta/2$.  We assume that $H_{\rm eff}=\Pi_-
H_{\rm eff} \Pi_-$ and ${\rm Spec}(H_{\rm eff}) \subseteq [a,b]$.
Let $0<\eps<\Delta$ and assume that for all $z \in D_r$, a disk of radius $r$ centered around
$z_0=\lambda_{0,\rm eff}$ with $\epsilon < r < \Delta_{\rm eff}-\epsilon$,
we have $\|\Sigma_-(z) - H_{\rm eff}\|\leq \eps.$
Let $\Pi_{0,\rm eff}$ be the projector onto the ground-space of $H_{\rm eff}$ with degeneracy $d$.
Let $\tilde{\Pi}_{\rm low}$ be the projector onto the $d$ lowest-lying eigenvectors of $\tilde{H}$.
Then we can bound \be ||\tilde{\Pi}_{\rm low}-\Pi_{0,\rm eff}|| \leq \frac{3\|V\|}{\lambda_+-(\lambda_{0,{\rm eff}}+\eps)}+
\frac{\epsilon (\lambda_{0,{\rm eff}}+r)}{r(r-\epsilon)}.\label{eq:groundbound}\ee
\label{lem:groundpert}
\end{lemma}

\proof{As in the proof of Theorem \ref{thm:perturbation}, we first prove that $\Pi_- \tilde{\Pi}_{\rm low}
\Pi_-$ is close to $\tilde{\Pi}_{\rm low}$. Then we show that
$\Pi_-\tilde{\Pi}_{\rm low} \Pi_-$ is close to the projector onto the ground
state of $H_{\rm eff}$, $\Pi_{0,\rm eff}$. For both parts we will use that due to the assumptions in the Lemma,
Theorem \ref{kkrtheorem} implies that
for all $i=0,\ldots, d-1$, $|\lambda_i(H_{\rm eff})-\lambda_i(\tilde{H})| \leq \epsilon$.
Let $\lambda_{0,\rm eff}$ be the lowest (degenerate) eigenvalue of $H_{\rm eff}$. We can first bound \be
||\tilde{\Pi}_{\rm low}-\Pi_- \tilde{\Pi}_{\rm low} \Pi_-|| \leq 3 ||\Pi_+
\tilde{\Pi}_{\rm low}||. \ee
As before we bound $||\Pi_+ H \tilde{\Pi}_{\rm low}||$ in two different directions:
\be
(\epsilon+\lambda_{0,{\rm eff}}) ||\Pi_+ \tilde{\Pi}_{\rm low}||+||V||\geq ||\Pi_+ H \tilde{\Pi}_{\rm low}|| \geq
\lambda_+||\Pi_+ \tilde{\Pi}_{\rm low}||.
\ee
These inequalities together with the previous equation give us the
first bound
\be ||\tilde{\Pi}_{\rm low}-\Pi_- \tilde{\Pi}_{\rm low} \Pi_-|| \leq \frac{3\|V\|}{\lambda_+-(\lambda_{0,{\rm eff}}+\eps)} \ee

In order to prove the other part we draw a circular contour $C$ of radius $r$ centered around
$z_0=\lambda_{0,\rm eff}$ with $\epsilon < r < \Delta_{\rm eff}-\epsilon$ such that it encloses only the lowest
$d$ eigenvalues of $\tilde{H}$. We will choose $r$ such that $\Delta_{\rm eff}-r > r$ or $z_0+r$ is closer to
$\lambda_{0,{\rm eff}}$ than to $\lambda_{1,{\rm eff}}$. We have \be \Pi_- \tilde{\Pi}_{\rm low}
\Pi_- = \frac{1}{2\pi i}\oint_C \Pi_-\tilde G(z)\Pi_- .\, \ee We use
that $||\Sigma_-(z)-H_{\rm eff}|| \leq \epsilon$ for $z \in C$ and
bound \be \sup_{z\in C}\|(zI_--\Sigma_-(z))^{-1} - (zI_--H_{\rm
eff})^{-1}\| \leq \frac{\epsilon}{r (r-\epsilon)}, \ee
using $||(zI_{-}-H_{\rm eff})^{-1}|| \leq r^{-1}$ for $z \in C$.
It follows that
\bea ||\Pi_- \tilde{\Pi}_{\rm low}\Pi_- -\Pi_{0,\rm eff}|| & \leq &
\frac{\epsilon (\lambda_{0,{\rm eff}}+r)}{r(r-\epsilon)}.\ \label{eq:part2new} \eea
}

\bibliographystyle{hunsrt}

\end{document}